\documentclass[prd,reprint,onecolumn,notitlepage,preprintnumbers,showpacs,superscriptaddress]{revtex4}
\usepackage{wrapfig}
\usepackage{verbatim}
\usepackage{amsmath, multirow, amssymb, wasysym, gensymb}
\usepackage{graphicx}
\usepackage{amsfonts}
\usepackage{graphicx}
\usepackage{graphics}
\usepackage[ngerman, english]{babel}
\usepackage{float}
\usepackage{youngtab}
\usepackage{subfigure}
\usepackage[utf8]{inputenc}
\usepackage[usenames,dvipsnames]{xcolor}
\usepackage[normalem]{ulem}

\input{tcilatex}
\begin{document}

\title{Higgs $\rightarrow $ $\mu \tau $ as an indication for $S_{4}$ flavor symmetry}
\author{Miguel D. Campos}
\affiliation{Max-Planck-Institut f\"ur Kernphysik, Saupfercheckweg 1, 69117 Heidelberg,
Germany}
\author{A. E. C\'{a}rcamo Hern\'{a}ndez}
\affiliation{{\ Universidad T\'ecnica Federico Santa Mar\'{\i}a and Centro Cient\'{\i}%
fico-Tecnol\'ogico de Valpara\'{\i}so}\\
Casilla 110-V, Valpara\'{\i}so, Chile}
\author{H. P\"as}
\affiliation{Fakult\"at f\"ur Physik, Technische Universit\"at Dortmund, 44221 Dortmund,
Germany}
\author{E. Schumacher}
\affiliation{Fakult\"at f\"ur Physik, Technische Universit\"at Dortmund, 44221 Dortmund,
Germany}

\begin{abstract}
\noindent Lepton flavor violating Higgs decays can arise in flavor symmetry
models where the Higgs sector is responsible for both the electroweak and
the flavor symmetry breaking. Here we advocate an $S_4$
three-Higgs-doublet model where tightly constrained flavor changing neutral
currents are suppressed by a remnant $Z_3$ symmetry. A small breaking of
this $Z_3$ symmetry can explain the $2.4\,\sigma$ excess of Higgs decay
final states with a $\mu \tau $ topology reported recently by CMS if the new neutral scalars are light.
The model also predicts
sizable rates for lepton flavor violating Higgs decays in the $e\tau $ and $%
e\mu $ channels because of the unifying $S_4$ flavor symmetry.
\end{abstract}

\pacs{11.30.Hv, 14.60.-z, 14.80.Ec}
\maketitle
\preprint{DO-TH 14/18}






\section{Introduction}

\label{s:intro}

Lepton flavor violating (LFV) Higgs decays have been advocated as a
harbinger of flavor symmetries explaining the large amount of lepton flavor
mixing \cite%
{Bhattacharyya:2010hp,Cao:2011df,Bhattacharyya:2012ze,Bhattacharyya:2012pi}. 
Indeed, substantial LFV Higgs couplings can arise quite
naturally in such models as a consequence of the maximal atmospheric $%
\mu-\tau$ mixing in the Pontecorvo-Maki-Nakagawa-Sakata (PMNS) matrix. To manifest itself in the physical mass
basis, a misalignment of the Higgs doublets, typically utilized to yield a
realistic symmetry breaking pattern, is necessary. While the scalar sector
in \cite%
{Bhattacharyya:2010hp,Cao:2011df,Bhattacharyya:2012ze,Bhattacharyya:2012pi}
decomposes into a Standard Model (SM)-like Higgs doublet and new exotic
scalars experiencing LFV decays, in the following we present an $S_4$ flavor
model where these states mix, resulting in sizable LFV decays of the SM-like
Higgs boson. This is particularly interesting after the recent report by the
CMS Collaboration of a $2.4\,\sigma$ anomaly in the $h\rightarrow \mu\tau$
channel with a best fit of Br$(h\rightarrow\mu\tau) \approx 0.84 \%$ \cite%
{Khachatryan:2015kon}, which as a possible hint of new physics beyond the SM, 
has drawn some attention \cite{Dery:2014kxa,Lee:2014rba,Celis:2014roa,Sierra:2014nqa,Heeck:2014qea,Dorsner:2015mja,Crivellin:2015lwa}.

Notably, the discrete group $S_4$ has been shown to be
the most natural flavor symmetry of the tribimaximal (TBM) mixing scheme in
the leptonic sector, with purely group theoretical arguments \cite%
{Hall:2013yha,Bazzocchi:2012st,Lam:2008sh}, as well as in explicit flavor
models \cite{Bazzocchi:2008ej,Ishimori:2011mt,Morisi:2010rk}. 
Furthermore, together with the groups $A_4$ and $\Delta(27)$, the $S_4$ group
is the smallest group containing an irreducible triplet representation that
can accommodate the three fermion families of the SM.
Nonsupersymmetric $S_4$ models based on TBM mixing but accommodating a
large $\theta_{13}$ have been discussed e.g. in \cite%
{Krishnan:2012me,Krishnan:2012sb,Mohapatra:2012tb}. To evade bounds from the
tightly constrained radiative decays $l_{\alpha} \rightarrow l_{\beta}
\gamma $, we consider the special case in which $S_4$ is broken down to a
residual $Z_3$ subgroup (see \cite{Cao:2011df} in the context of the
symmetry groups $A_4$, $T_7$, and $\Delta(27)$, referred to as lepton flavor
triality or LFT for short). This discrete $Z_3$ symmetry is obtained when
scalar doublets in the irreducible triplet representation $\mathbf{3^{\prime
}}$ of $S_4$ assume the specific vacuum alignment $(1,1,1)$ \cite%
{Bazzocchi:2008ej}.
If only the charged lepton sector is considered, the distinct $Z_3$ quantum numbers prevent any mixing of the physical scalars in this model. The $Z_3$ symmetry is, however, broken to some degree by perturbations arising from
other scalar $S_4$ triplets required to extend the model to quarks and neutrinos.
These perturbations enable mixing between the scalars and consequently lead to
LFV Higgs decays.

The measurement of $h\rightarrow \mu\tau$ can be translated into a bound on
the combination of Yukawa couplings $|y_{\mu\tau}|^2+|y_{\tau\mu}|^2$ \cite%
{Dorsner:2015mja}. These couplings directly affect predictions for
constrained LFV processes such as $\tau \rightarrow \mu \gamma$ or $\tau
\rightarrow 3 \mu$ which can be used to set bounds on the scalar masses in our 3HDM.

The paper is structured as follows. In Sec. \ref{s:model} we develop a model 
with LFT that leads to realistic fermion masses and mixings.
In Sec. \ref{s:perturb} we analyze the breaking of the $Z_3$ group associated 
with LFT and its consequences for Higgs decays. In particular we focus
on the $h\rightarrow \mu\tau$ channel explaining the recent $2.4\,\sigma$
anomaly and summarize predictions of our model for other rare LFV processes.
The $h\rightarrow \gamma \gamma$ rate is considered in Sec. \ref{s:hgg} and a summary is given in Sec.~\ref{s:summ}.

\section{The Model}

\label{s:model}

\subsection{Charged leptons}

In the following we use the charged lepton sector
as a starting point to introduce the relevant scalar content. Some of the additional
scalars needed to accommodate quarks and neutrinos have an effect on the lepton phenomenology, which we can
exploit to explain the excess in $h \rightarrow \mu\tau$ reported by CMS, as
will be discussed in Sec. \ref{s:perturb}. \newline
The particle assignments relevant for the charged lepton sector in the
notation $(S_{4},SU(2),Z_{12})$ are 
\begin{equation}  \label{e:assign}
\begin{aligned} L=(L_e,L_{\mu},L_{\tau})~:~(3^{\prime },2,1)\, ,\quad &
\tau_R~ :~(1,1,-i)\, ,\quad (e_R, \mu_R)~:~(2,1,e^{\frac{7 i\pi}{6}})\, , \\
\phi=(\phi_1,\phi_2,\phi_3)~:~(3^{\prime },2,1)\, , \quad &
\eta_1~:~(1,1,e^{\frac{i\pi}{6}}), \quad \eta_2~:~(1^{\prime
},1,e^{\frac{i\pi}{6}})\,. \end{aligned}
\end{equation}
The scalar fields $\phi _{j}$ ($j=1,2,3$) and the left-handed fermion $SU(2)$
doublets are both assigned to an $S_{4}$ triplet representation $\mathbf{%
3^{\prime }}$. The former break the electroweak (EW) symmetry of the SM by
spontaneously acquiring VEVs at the EW scale. Such models are usually
referred to as three-Higgs-doublet models (3HDMs) and have been extensively
analyzed in the past, e.g., in \cite{Felipe:2013ie,Keus:2013hya}.

To describe the hierarchy among the charged fermion masses, we introduce two
EW scalar singlets, $\eta _{1}$ and $\eta _{2}$. For the VEVs of $\eta _{1,2}$ we assume $v_{\eta _{1,2}}=\lambda \Lambda$
with $\lambda \approx 0.22$, where $\Lambda$ is a high scale defining the
breakdown of our effective theory.
By choosing suitable $Z_{12}$ charges we ensure that the tau and muon mass 
arise from seven and nine dimensional Yukawa terms, respectively. The 
smallness of the electron mass on the other hand is explained
by the destructive interference between the contributions
coming from the nine dimensional Yukawa operators and by a small breaking of
the universality of the corresponding Yukawa couplings.
The $Z_{12}$, therefore, functions as a
Froggatt--Nielson-like symmetry by creating mass hierarchies in the lepton
and quark sector with $\mathcal{O}(1)$ Yukawa couplings \cite{Wang:2011ub}.

As a consequence of the particle assignments given in Eq. (\ref{e:assign})
the scalar potential $V(\phi )$ involving only the $\phi $ field is the
general $S_{4}$-invariant scalar potential of a 3HDM \cite{Keus:2013hya}.
Since the singlet scalars 
are assumed to be very heavy, the mixing between $\phi_i$ and $\eta_{1,2}$
is suppressed. For simplicity we also assume a CP-conserving scalar
potential with only real couplings as done in, e.g., \cite{Machado:2010uc}.
The renormalizable low-energy scalar potential is%
\begin{align}
V(\phi )=& -\mu _{1}^{2}\sum_{i=1}^{3}\phi _{i}^{\dagger }\phi _{i}+\alpha
\left (\sum_{i=1}^{3}\phi _{i}^{\dagger }\phi _{i}\right
)^{2}+\sum_{i,j=1,i\neq j}^{3}(\beta (\phi _{i}^{\dagger }\phi _{i})(\phi
_{j}^{\dagger }\phi _{j})+\gamma |\phi _{i}^{\dagger }\phi _{j}|^{2}+\delta
(\phi _{i}^{\dagger }\phi _{j})^{2})  \label{eq:V1}
\end{align}

with the low-energy scalar content given by 
\begin{equation}  \label{eq:scalars}
\begin{aligned} \phi_j = \begin{bmatrix} {\phi_j^{+}}, \quad
{\frac{1}{\sqrt{2}}\left(\frac{v}{\sqrt{3}} + \phi^0_{jR} +
i\phi^0_{jI}\right) }\end{bmatrix}.\end{aligned}
\end{equation}
It includes three CP-even neutral scalars $\phi _{jR}^{0}$ ($j=1,2,3$), and three CP-odd neutral scalars $\phi _{jI}^{0}$, as well as three complex charged scalars ($%
\phi _{j}^{+}$), of which three degrees of freedom are absorbed by the $%
W^{\pm }$ and $Z$ gauge boson masses. 
The corresponding physical mass spectrum reads 
\begin{equation}
\begin{array}{rclcrclcrcl}
\label{eq:phimass} m^2_{\phi^0_{(a,b)R}} & = & -\tfrac{v^2}{3}\kappa, & 
\hspace{0.25cm} & m^2_{\phi^0_{cR}} & = & \tfrac{v^2}{3}(3\alpha+2 \kappa),
& \hspace{0.25cm} & m^2_{\phi^0_{(a,b)I}} & = & -v^2\delta, \\ 
m^2_{\phi^{\pm}_{a,b}} & = & -v^2(\kappa-\beta), & \hspace{0.25cm} & 
m^2_{\phi^0_{cI}} & = & 0, & \hspace{0.25cm} & m^2_{\phi^{\pm}_c} & = & 0,%
\end{array}%
\end{equation}
with $\kappa := \beta + \gamma + \delta$.

The mass eigenstates $\phi_{a,b,c}$ are given by the following linear
combinations of the $S_4$ basis scalars $\phi_{1,2,3}$: 
\begin{equation}
\begin{array}{rclcrclcrcl}
\label{eq:physsc}\phi_a & = & \frac{1}{\sqrt{2}}(\phi_3 - \phi_2), & \hspace{%
0.25cm} & \phi_b & = & \frac{1}{\sqrt{6}} (2\phi_1 - \phi_2 - \phi_3), & 
\hspace{0.25cm} & \phi_c & = & \frac{1}{\sqrt{3}} (\phi_1 + \phi_2 + \phi_3).%
\end{array}%
\end{equation}
These equations hold for the charged, CP-even and CP-odd components of $\phi$%
, and they imply that 
\begin{align}  \label{e:vevs}
\langle \phi_a \rangle = \langle \phi_b \rangle = 0 \quad \text{and} \quad
\langle \phi_c \rangle &= v.
\end{align}
The fact that $\phi_c$ is the only mass eigenstate from the $S_4$ triplet
acquiring a VEV proves essential for LFT. Along with the degeneracy of the
scalars $\phi _{(a,b)R}^{0}$ (cf. Eq. (\ref{eq:phimass})), it
suggests that $\phi _{cR}^{0}$ can be identified as the SM Higgs particle
found at the LHC with a mass of approximately 125 GeV.

Using Eq. (\ref{eq:physsc}) we obtain the three- and four-point vertices of
the physical scalars and the gauge bosons $W^{\pm}, Z^0$ and $A$ from the
kinetic terms of the Lagrangian. In the case of the three-point
interactions, the only induced decay channels are $\phi^0_{cR}\to W^+ W^-$
and $\phi^0_{cR}\to (Z^0)^2$. Hence, the SM-like Higgs $\phi_{cR}^0$ is the
only physical scalar giving masses to the SM gauge bosons with SM couplings.
As a consequence, the masses of the neutral scalars $\phi^0_{(a,b)R}$ are
not constrained by the usual Higgs searches performed in the LEP and LHC
experiments \cite{ATLAS-CONF-2012-019,CMS-PAS-HIG-12-008}.

The relevant $S_{4}\otimes Z_{12}$-invariant Yukawa terms for the charged
leptons 
\begin{equation}
\mathcal{L}\supset y_{1}\left[ L\phi \right] _{1}\tau _{R}\frac{\eta
_{1}^{3}+\varepsilon _{0}\eta _{1}\eta _{2}^{2}}{\Lambda ^{3}}+y_{2}\left[
L\phi \right] _{2}\left( 
\begin{array}{c}
e_{R} \\ 
\mu _{R}%
\end{array}%
\right) \frac{\eta _{1}^{5}+\varepsilon _{1}\eta _{1}^{3}\eta
_{2}^{2}+\varepsilon _{2}\eta _{1}\eta _{2}^{4}}{\Lambda ^{5}}+y_{3}\left[
L\phi \right] _{2}\left( 
\begin{array}{c}
e_{R} \\ 
\mu _{R}%
\end{array}%
\right) \frac{\eta _{2}^{5}+\varepsilon _{3}\eta _{2}^{3}\eta
_{1}^{2}+\varepsilon _{4}\eta _{2}\eta _{1}^{4}}{\Lambda ^{5}}  \label{e:ll}
\end{equation}

lead to the following mass matrix for charged leptons: 
\begin{align}
M_{l}&=\frac{1}{\sqrt{2}}\left( 
\begin{array}{ccc}
v_{1}\left( \widetilde{y}_{2}-\widetilde{y}_{3}\right) \lambda ^{5} & 
v_{1}\left( \widetilde{y}_{2}+\widetilde{y}_{3}\right) \lambda ^{5} & v_{1}%
\widetilde{y}_{1}\lambda ^{3} \\ 
\label{e:vlL} v_{2}\omega ^{2}\left( \widetilde{y}_{2}-\widetilde{y}%
_{3}\right) \lambda ^{5} & v_{2}\omega \left( \widetilde{y}_{2}+\widetilde{y}%
_{3}\right) \lambda ^{5} & v_{2}\widetilde{y}_{1}\lambda ^{3} \\ 
v_{3}\omega \left( \widetilde{y}_{2}-\widetilde{y}_{3}\right) \lambda ^{5} & 
v_{3}\omega ^{2}\left( \widetilde{y}_{2}+\widetilde{y}_{3}\right) \lambda
^{5} & v_{3}\widetilde{y}_{1}\lambda ^{3}%
\end{array}%
\right) \\
&=v\frac{1}{\sqrt{6}}\left( 
\begin{array}{ccc}
1 & 1 & 1 \\ 
\omega ^{2} & \omega & 1 \\ 
\omega & \omega ^{2} & 1%
\end{array}%
\right) \text{diag}(\left( \widetilde{y}_{2}-\widetilde{y}_{3}\right)
\lambda ^{5},\left( \widetilde{y}_{2}+\widetilde{y}_{3}\right) \lambda ^{5},%
\widetilde{y}_{1}\lambda ^{3})
\end{align}%
with $v_{1}=v_{2}=v_{3}=\tfrac{v}{\sqrt{3}}$, $\omega =e^{2i\pi /3}$ and $%
\widetilde{y}_{1}$, $\widetilde{y}_{2}$, $\widetilde{y}_{3}$ given by 
\begin{equation}
\widetilde{y}_{1}=\left( 1+\varepsilon _{0}\right) y_{1},\hspace{0.2cm}%
\hspace{0.2cm}\hspace{0.2cm}\hspace{0.2cm}\hspace{0.2cm}\hspace{0.2cm}%
\hspace{0.2cm}\widetilde{y}_{2}=\left( 1+\varepsilon _{1}+\varepsilon
_{2}\right) y_{2},\hspace{0.2cm}\hspace{0.2cm}\hspace{0.2cm}\hspace{0.2cm}%
\hspace{0.2cm}\hspace{0.2cm}\hspace{0.2cm}\widetilde{y}_{3}=\left(
1+\varepsilon _{3}+\varepsilon _{4}\right) y_{3},
\end{equation}%
where the dimensionless couplings $y_{1}$, $y_{2}$, $y_{3}$, $\varepsilon
_{0}$, $\varepsilon _{1}$, $\varepsilon _{2}$, $\varepsilon _{3}$ and $%
\varepsilon _{4}$ are $\mathcal{O}(1)$ parameters. Consequently the charged
lepton masses are 
\begin{equation}
m_{e}=\left( \widetilde{y}_{2}-\widetilde{y}_{3}\right) \lambda ^{5}\frac{v}{%
\sqrt{2}},\hspace{0.2cm}\hspace{0.2cm}\hspace{0.2cm}\hspace{0.2cm}\hspace{%
0.2cm}\hspace{0.2cm}\hspace{0.2cm}m_{\mu }=\left( \widetilde{y}_{2}+%
\widetilde{y}_{3}\right) \lambda ^{5}\frac{v}{\sqrt{2}},\hspace{0.2cm}%
\hspace{0.2cm}\hspace{0.2cm}\hspace{0.2cm}\hspace{0.2cm}\hspace{0.2cm}%
\hspace{0.2cm}m_{\tau }=\widetilde{y}_{1}\lambda ^{3}\frac{v}{\sqrt{2}}.
\end{equation}%
The mass hierarchy $m_{e},m_{\mu }\ll m_{\tau }$, therefore, is a natural
consequence of our model. 

Equation (\ref{e:vlL}) shows a characteristic feature of LFT that the mass
basis of the charged leptons coincides with the $Z_3$ basis; i.e.,
the charged lepton fields can be identified as $Z_3$ eigenstates $e \sim
1,\mu \sim \omega^2,\tau \sim \omega$.


The structures for the couplings of the charged leptons to the physical
scalars $\phi^0_{aR}, \phi^0_{bR}$ and $\phi^0_{cR}$ are given by 
\begin{align}  \label{e:yc}
Y_{(a)b} =(i)\frac{1}{v\sqrt{2}}\left( 
\begin{array}{ccc}
0 & m_{\mu} \omega^2 & (-)m_{\tau} \omega \\ 
(-)m_{e} \omega & 0 & m_{\tau}\omega^2 \\ 
m_e \omega^2 & (-)m_{\mu}\omega & 0%
\end{array}
\right),\quad Y_c =\frac{1}{v}\left( 
\begin{array}{ccc}
m_e & 0 & 0 \\ 
0 & m_{\mu} & 0 \\ 
0 & 0 & m_{\tau}%
\end{array}
\right),
\end{align}
where the factors in parentheses apply only to the structure of $Y_a$. 
As can be seen from Eq. (\ref{e:yc}), $\phi_c$ only couples diagonally to
the charged leptons and, hence, should be uncharged under $Z_3$. This can be
understood by expressing the fields $\phi_{a,b,c}$ in terms of the $Z_3$
eigenstates 
\begin{align}
\left( 
\begin{array}{c}
\phi_x \\ 
\phi_y \\ 
\phi_z%
\end{array}
\right)=\frac{1}{\sqrt{3}} \left( 
\begin{array}{ccc}
1 & 1 & 1 \\ 
1 & \omega^2 & \omega \\ 
1 & \omega & \omega^2%
\end{array}
\right) \left( 
\begin{array}{c}
\phi_1 \\ 
\phi_2 \\ 
\phi_3%
\end{array}
\right) \qquad \text{with} \quad \phi_{x,y,z} \sim 1, \omega,\omega^2
\end{align}
and consequently realizing that 
\begin{align}
\phi_c \equiv \phi_x, \qquad \phi_b \equiv \frac{1}{\sqrt{2}}%
(\phi_y+\phi_z), \qquad \phi_a \equiv \frac{1}{\sqrt{2}}(\phi_y-\phi_z).
\end{align}
Considering that $\phi_c$ is the only mass eigenstate from the $\phi$ $S_4$
triplet that acquires a VEV (cf. Eq. (\ref{e:vevs})), $Z_3$ remains
unbroken at this point. 


\subsection{Neutrino sector}

To generate the neutrino masses via a type I seesaw mechanism, we extend the
SM particle content by two heavy Majorana neutrinos $N_{1R}$ and $N_{2R}$ as
well as four $S_{4}$ triplets scalar fields $\chi $, $\xi $, $\sigma $\ and $%
\zeta $, which are singlets under $SU(2)$. Additionally we employ two $Z_{2}$
symmetries to enforce a specific mass pattern and decouple the scalars from
interactions with the other fermion sectors. The corresponding $S_{4}\otimes
Z_{2}\otimes Z_{2}^{\prime }\otimes Z_{12}$ assignments are 
\begin{equation}
\begin{aligned}& N_{1R}~:~(1,-1,1,1)\,,\quad N_{2R}~:~(1,-1,-1,1), \\ \chi :
~(3^{\prime },-1,1,1)&\,,\quad \xi ~:~(3,-1,1,1)\,,\quad \sigma :~(3^{\prime
},-1,-1,1)\,,\quad \zeta :~(3,-1,-1,1). \end{aligned}  \label{e:numasses}
\end{equation}%
Therefore the relevant $S_{4}\otimes Z_{2}\otimes Z_{2}^{\prime }\otimes
Z_{12}$ -invariant neutrino Yukawa terms read 
\begin{eqnarray}
\mathcal{L} &\supset &y_{1}^{\left( \nu \right) }\left[ L\phi \right]
_{3^{\prime }}N_{1R}~\frac{\chi }{\Lambda }+y_{2}^{\left( \nu \right) }\left[
L\phi \right] _{3}N_{1R}~\frac{\xi }{\Lambda }+y_{3}^{\left( \nu \right) }%
\left[ L\phi \right] _{3^{\prime }}N_{2R}~\frac{\sigma }{\Lambda }%
+y_{4}^{\left( \nu \right) }\left[ L\phi \right] _{3}N_{2R}~\frac{\zeta }{%
\Lambda }+\frac{y_{5}^{\left( \nu \right) }}{\Lambda }L\left[ \phi \chi %
\right] _{3^{\prime }}N_{1R}~  \notag \\
&&+\frac{y_{6}^{\left( \nu \right) }}{\Lambda }L\left[ \phi \xi \right]
_{3^{\prime }}N_{1R}+\frac{y_{7}^{\left( \nu \right) }}{\Lambda }L\left[
\phi \sigma \right] _{3^{\prime }}N_{2R}+\frac{y_{8}^{\left( \nu \right) }}{%
\Lambda }L\left[ \phi \zeta \right] _{3^{\prime }}N_{2R}+M_{1}\overline{N}%
_{1R}N_{1R}^{c}+M_{2}\overline{N}_{2R}N_{2R}^{c}.
\end{eqnarray}%
With the VEV patterns of the scalar fields $\chi $, $\xi $, $\sigma $\ and $%
\zeta $ 
\begin{equation}  \label{e:vevsnu}
\left\langle \chi \right\rangle =v_{\chi }\left( 1,0,0\right) \,,\quad
\left\langle \xi \right\rangle =v_{\xi }\left( 1,0,0\right) \,,\quad
\left\langle \sigma \right\rangle =v_{\sigma }\left( 0,i,0\right) \,,\quad
\left\langle \zeta \right\rangle =v_{\zeta }\left( 0,1,0\right) \,\,
\end{equation}%
and the simplification 
\begin{equation}
y_{1}^{\left( \nu \right) }=y_{2}^{\left( \nu \right) }=y_{5}^{\left( \nu
\right) }=y_{6}^{\left( \nu \right) }=y^{\left( \nu \right) }\,,
\end{equation}%
it follows that the full $5\times 5$ neutrino mass matrix is 
\begin{equation}
M_{L}^{\left( \nu \right) }=\left( 
\begin{array}{cc}
0_{3\times 3} & M_{\nu }^{D} \\ 
\left( M_{\nu }^{D}\right) ^{T} & M_{R}%
\end{array}%
\right) ,
\end{equation}%
where 
\begin{eqnarray}
M_{\nu }^{D} &=&\left( 
\begin{array}{cc}
0 & \lambda \left[ \left( y_{4}^{\left( \nu \right) }+y_{6}^{\left( \nu
\right) }\right) -i\left( y_{3}^{\left( \nu \right) }+y_{7}^{\left( \nu
\right) }\right) \right] \\ 
4\lambda y^{\left( \nu \right) } & 0 \\ 
0 & \lambda \left[ \left( y_{4}^{\left( \nu \right) }+y_{6}^{\left( \nu
\right) }\right) +i\left( y_{3}^{\left( \nu \right) }+y_{7}^{\left( \nu
\right) }\right) \right]%
\end{array}%
\right) \frac{v}{\sqrt{3}}=\left( 
\begin{array}{cc}
0 & ae^{i\tau } \\ 
b & 0 \\ 
0 & ae^{-i\tau }%
\end{array}%
\right) \frac{v}{\sqrt{3}},\hspace{1cm}M_{R}=\left( 
\begin{array}{cc}
M_{1} & 0 \\ 
0 & M_{2}%
\end{array}%
\right) ,  \notag \\
a &=&\lambda \sqrt{\left( y_{4}^{\left( \nu \right) }+y_{6}^{\left( \nu
\right) }\right) ^{2}+\left( y_{3}^{\left( \nu \right) }+y_{7}^{\left( \nu
\right) }\right) ^{2}},\hspace{1cm}\hspace{1cm}b=4\lambda y^{\left( \nu
\right) },\hspace{1cm}\hspace{1cm}\tan \tau =-\frac{y_{3}^{\left( \nu
\right) }+y_{7}^{\left( \nu \right) }}{y_{4}^{\left( \nu \right)
}+y_{6}^{\left( \nu \right) }}.
\end{eqnarray}

Since $\left( M_{R}\right) _{ii}\gg v$, the light neutrino mass matrix
arises from a type I seesaw mechanism and is given by: 
\begin{equation}
M_{L}^{\left( \nu \right) }=M_{\nu }^{D}M_{R}^{-1}\left( M_{\nu }^{D}\right)
^{T}=\left( 
\begin{array}{ccc}
a^{2}e^{2i\tau } & 0 & a^{2} \\ 
0 & b^{2}\frac{M_{2}}{M_{1}} & 0 \\ 
a^{2} & 0 & a^{2}e^{-2i\tau }%
\end{array}%
\right) \frac{v^{2}}{3M_{2}}=\left( 
\begin{array}{ccc}
Ae^{2i\tau } & 0 & A \\ 
0 & B & 0 \\ 
A & 0 & Ae^{-2i\tau }%
\end{array}%
\right)  \label{Mnu}
\end{equation}%
with 
\begin{equation}
A=a^{2}\frac{v^{2}}{3M_{2}},\hspace{0.2cm}\hspace{0.2cm}\hspace{0.2cm}%
\hspace{0.2cm}\hspace{0.2cm}\hspace{0.2cm}\hspace{0.2cm}B=b^{2}\frac{v^{2}}{%
3M_{1}}.
\end{equation}%
It is worth mentioning that the neutrino mass matrix depends only on three
effective parameters, $A,B$ and $\tau $, of which $A$ and $B$ are inverse
proportional to $M_{2}$ and $M_{1}$, respectively. Furthermore, the
smallness of the active neutrino masses arises from their inverse scaling with
the large Majorana neutrino masses as well as from the quadratic
dependence of the neutrino Yukawa couplings, which is a characteristic
feature of the type I seesaw mechanism. The right handed Majorana neutrinos obtain large masses
due to their Yukawa interactions with EW scalar singlets, which
acquire VEVs much larger than the electroweak scale. 

The squared mass matrix $M_{L}^{\left( \nu \right) }\left( M_{L}^{\left( \nu
\right) }\right) ^{\dagger }$ is diagonalized by a unitary rotation matrix $%
V_{\nu }$ as follows \cite%
{Hernandez:2013dta,Campos:2014lla,Hernandez:2015tna,Hernandez:2015cra}: 
\begin{equation}
V_{\nu }^{\dagger }M_{\nu }^{\left( 1\right) }\left( M_{\nu }^{\left(
1\right) }\right) ^{\dagger }V_{\nu }=\left( 
\begin{array}{ccc}
m_{1}^{2} & 0 & 0 \\ 
0 & m_{2}^{2} & 0 \\ 
0 & 0 & m_{3}^{2}%
\end{array}%
\right) ,\hspace{0.5cm}\mbox{with}\hspace{0.5cm}V_{\nu }=\left( 
\begin{array}{ccc}
\cos \psi & 0 & \sin \psi e^{-i\phi } \\ 
0 & 1 & 0 \\ 
-\sin \psi e^{i\phi } & 0 & \cos \psi%
\end{array}%
\right) ,\hspace{0.5cm}\psi =\pm \frac{\pi }{4},\hspace{0.5cm}\phi =-2\tau ,
\label{Vnu}
\end{equation}%
where $\psi =+\pi /4$ and $\psi =-\pi /4$ correspond to the normal (NH)
and inverted (IH) mass hierarchies, respectively. The light active neutrino
masses for NH and IH are: 
\begin{eqnarray}
\mbox{NH} &:&\psi =+\frac{\pi }{4}:\hspace{10mm}m_{\nu _{1}}=0,\hspace{10mm%
}m_{\nu _{2}}=B,\hspace{10mm}m_{\nu _{3}}=2\left\vert A\right\vert ,
\label{mass-spectrum-Inverted} \\[0.12in]
\mbox{IH} &:&\psi =-\frac{\pi }{4}:\hspace{10mm}m_{\nu _{1}}=2\left\vert
A\right\vert ,\hspace{8mm}m_{\nu _{2}}=B,\hspace{10mm}m_{\nu _{3}}=0.
\label{mass-spectrum-Normal}
\end{eqnarray}%
By combining Eqs. (\ref{e:vlL}) and (\ref{Vnu}) we obtain the
PMNS leptonic mixing matrix 
\begin{equation}
U=V_{lL}^{\dag }V_{\nu }=\left( 
\begin{array}{ccc}
\frac{\cos \psi }{\sqrt{3}}-\frac{e^{i\phi -\frac{2i\pi }{3}}\sin \psi }{%
\sqrt{3}} & \frac{e^{\frac{2i\pi }{3}}}{\sqrt{3}} & \frac{e^{-\frac{2i\pi }{3%
}}\cos \psi }{\sqrt{3}}+\frac{e^{-i\phi }\sin \psi }{\sqrt{3}} \\ 
&  &  \\ 
\frac{\cos \psi }{\sqrt{3}}-\frac{e^{i\phi +\frac{2i\pi }{3}}\sin \psi }{%
\sqrt{3}} & \frac{e^{-\frac{2i\pi }{3}}}{\sqrt{3}} & \frac{e^{\frac{2i\pi }{3%
}}\cos \psi }{\sqrt{3}}+\frac{e^{-i\phi }\sin \psi }{\sqrt{3}} \\ 
&  &  \\ 
\frac{\cos \psi }{\sqrt{3}}-\frac{e^{i\phi }\sin \psi }{\sqrt{3}} & \frac{1}{%
\sqrt{3}} & \frac{\cos \psi }{\sqrt{3}}+\frac{e^{-i\phi }\sin \psi }{\sqrt{3}%
}%
\end{array}%
\right) ,  \label{PMNS}
\end{equation}%
which only depends on a single parameter $\phi $, whereas the neutrino mass
squared splittings are determined by the parameters $A$ and $B$.

Furthermore, we find that the lepton mixing angles are given by: 
\begin{eqnarray}
&&\sin ^{2}\theta _{12}=\frac{\left\vert U_{e2}\right\vert ^{2}}{%
1-\left\vert U_{e3}\right\vert ^{2}}=\frac{2}{4\pm \left( \cos \phi -\sqrt{3}%
\sin \phi \right) },\hspace{20mm}  \label{theta-ij} \\[3mm]
&&\sin ^{2}\theta _{13}=\left\vert U_{e3}\right\vert ^{2}=\frac{1}{6}\left[
2\mp (\cos \phi -\sqrt{3}\sin \phi )\right] , \\
&&\sin ^{2}\theta _{23}=\frac{\left\vert U_{\mu 3}\right\vert ^{2}}{%
1-\left\vert U_{e3}\right\vert ^{2}}=\frac{2\mp (\cos \phi +\sqrt{3}\sin
\phi )}{4\pm \left( \cos \phi -\sqrt{3}\sin \phi \right) }.
\end{eqnarray}%
The Jarlskog invariant and the CP-violating phase are given by \cite%
{PDG-2014} 
\begin{equation}  \label{e:jnp}
J=\func{Im}\left( U_{e1}U_{\mu 2}U_{e2}^{\ast }U_{\mu 1}^{\ast }\right) =%
\frac{1}{6\sqrt{3}}\cos 2\psi ,\hspace{2cm}\sin \delta =\frac{8J}{\cos
\theta _{13}\sin 2\theta _{12}\sin 2\theta _{23}\sin 2\theta _{13}}.
\end{equation}%
Since $\psi =\pm \frac{\pi }{4}$, we predict $J=0$ and $\delta =0$, which
corresponds to CP conservation in neutrino oscillations.

In what follows we adjust the three free effective parameters $\phi $, $A$
and $B$ of the light neutrino sector to accommodate the experimental values
of three leptonic mixing parameters and two neutrino mass squared
splittings, reported in Tables \mbox{\ref{NH} and \ref{IH}} for NH and IH,
respectively. We fit the $\phi $ parameter to adjust the values of the
leptonic mixing parameters $\sin ^{2}\theta _{ij}$, whereas $A$ and $B$ are
given by 
\begin{eqnarray}
&&\mbox{NH}:\ m_{\nu _{1}}=0,\ \ \ m_{\nu _{2}}=B=\sqrt{\Delta m_{21}^{2}}%
\approx 9\,\mbox{meV},\ \ \ m_{\nu _{3}}=2\left\vert A\right\vert =\sqrt{%
\Delta m_{31}^{2}}\approx 50\,\mbox{meV};  \label{AB-Delta-IH} \\[0.12in]
&&\mbox{IH}\hspace{2mm}:\ m_{\nu _{2}}=B=\sqrt{\Delta m_{21}^{2}+\Delta
m_{13}^{2}}\approx 50\,\mbox{meV},\ \ \ \ \ m_{\nu _{1}}=2\left\vert
A\right\vert =\sqrt{\Delta m_{13}^{2}}\approx 49\,\mbox{meV},\ \ \ m_{\nu
_{3}}=0,  \label{AB-Delta-NH}
\end{eqnarray}%
resulting from Eqs. (\ref{mass-spectrum-Inverted}) and (\ref%
{mass-spectrum-Normal}), the definition $\Delta
m_{ij}^{2}=m_{i}^{2}-m_{j}^{2}$, and the best-fit values of $\Delta
m_{ij}^{2}$ from Tables \ref{NH} and \ref{IH} for NH and IH, respectively.

By varying $\phi$ we obtain the following best-fit result: 
\begin{eqnarray}
&&\mbox{NH}\ :\ \phi =-0.453\pi ,\ \ \ \sin ^{2}\theta _{12}\approx 0.34,\ \
\ \sin ^{2}\theta _{23}\approx 0.61,\ \ \ \sin ^{2}\theta _{13}\approx
0.0232;  \label{parameter-fit-IH} \\[0.12in]
&&\mbox{IH}\hspace{2.5mm}:\ \phi =\ \ 0.546\,\pi ,\ \ \ \ \ \sin ^{2}\theta
_{12}\approx 0.34,\ \ \ \sin ^{2}\theta _{23}\approx 0.61,\ \ \ \ \,\sin
^{2}\theta _{13}\approx 0.024.  \label{parameter-fit-NH}
\end{eqnarray}

Comparing Eqs. (\ref{parameter-fit-IH}) and (\ref{parameter-fit-NH}) with
Tables \ref{NH} and \ref{IH}, we obtain $\sin ^{2}\theta _{13}$ and $\sin
^{2}\theta _{23}$ in excellent agreement with the experimental data, for
both mass hierarchies, whereas $\sin ^{2}\theta _{12}$ deviates $2\sigma $
away from its best-fit values. Consequently, our predictions for the
neutrino mass squared splittings and leptonic mixing parameters 
are in very good agreement with the experimental data on neutrino
oscillations. Furthermore, our model predicts the absence of CP violation in
neutrino oscillations.%

\begin{table}[tbh]
\begin{tabular}{|c|c|c|c|c|c|}
\hline
Parameter & $\Delta m_{21}^{2}$($10^{-5}$eV$^2$) & $\Delta m_{31}^{2}$($%
10^{-3}$eV$^2$) & $\left( \sin ^{2}\theta _{12}\right) _{\exp }$ & $\left(
\sin ^{2}\theta _{23}\right) _{\exp }$ & $\left( \sin ^{2}\theta
_{13}\right) _{\exp }$ \\ \hline
Best fit & $7.60$ & $2.48$ & $0.323$ & $0.567$ & $0.0234$ \\ \hline
$1\sigma $ range & $7.42-7.79$ & $2.41-2.53$ & $0.307-0.339$ & $0.439-0.599$
& $0.0214-0.0254$ \\ \hline
$2\sigma $ range & $7.26-7.99$ & $2.35-2.59$ & $0.292-0.357$ & $0.413-0.623$
& $0.0195-0.0274$ \\ \hline
$3\sigma $ range & $7.11-8.11$ & $2.30-2.65$ & $0.278-0.375$ & $0.392-0.643$
& $0.0183-0.0297$ \\ \hline
\end{tabular}%
\caption{Range for experimental values of neutrino mass squared splittings
and leptonic mixing parameters, taken from Ref. \protect\cite{Forero:2014bxa}%
, for the case of normal hierarchy.}
\label{NH}
\end{table}
\begin{table}[tbh]
\begin{tabular}{|c|c|c|c|c|c|}
\hline
Parameter & $\Delta m_{21}^{2}$($10^{-5}$eV$^{2}$) & $\Delta m_{13}^{2}$($%
10^{-3}$eV$^{2}$) & $\left( \sin ^{2}\theta _{12}\right) _{\exp }$ & $\left(
\sin ^{2}\theta _{23}\right) _{\exp }$ & $\left( \sin ^{2}\theta
_{13}\right) _{\exp }$ \\ \hline
Best fit & $7.60$ & $2.38$ & $0.323$ & $0.573$ & $0.0240$ \\ \hline
$1\sigma $ range & $7.42-7.79$ & $2.32-2.43$ & $0.307-0.339$ & $0.530-0.598$
& $0.0221-0.0259$ \\ \hline
$2\sigma $ range & $7.26-7.99$ & $2.26-2.48$ & $0.292-0.357$ & $0.432-0.621$
& $0.0202-0.0278$ \\ \hline
$3\sigma $ range & $7.11-8.11$ & $2.20-2.54$ & $0.278-0.375$ & $0.403-0.640$
& $0.0183-0.0297$ \\ \hline
\end{tabular}%
\caption{Range for experimental values of neutrino mass squared splittings
and leptonic mixing parameters, taken from Ref. \protect\cite{Forero:2014bxa}%
, for the case of inverted hierarchy.}
\label{IH}
\end{table}

\subsection{Quark sector}

To obtain realistic quark masses and mixings 
we add SM scalar singlets, i.e., two $S_{4}$ triplets, $\rho $ and $\varphi$,
and three $S_{4}$ singlets $\Omega _{1}$, $\Omega _{2}$ and $\Omega _{3}$.
Again we use a $Z_2^{\prime \prime}$ symmetry to decouple these scalars from
the other fermion sectors, whereas a $Z_6$ symmetry accounts for the top and
bottom mass hierarchy. The $S_{4}\otimes Z_{2}^{\prime \prime }\otimes
Z_{6}\otimes Z_{12}$ assignments are: 
\begin{eqnarray}
t_{R}&:&~(1,1,1,1),\quad c_{R}:~(1,1,1,1),\quad u_{R}:~(1,1,1,-1),\quad
b_{R}:~(1,1,-1,1),\quad s_{R}:~(1,1,-1,1),\quad d_{R}:~(1,1,-1,-1),  \notag
\\
Q&:&~(3^{\prime },-1,1,1),\ \ \rho:~(3^{\prime },-1,1,1),\ \ \varphi
:~(3,-1,1,1),\ \ \Omega _{1}:~(1,1,1,i),\ \ \Omega _{2}:~(1,-1,1,\omega ^{%
\frac{1}{2}}),\ \ \Omega _{3}:~(1,1,\omega ^{\frac{1}{2}},1)  \label{qa}
\end{eqnarray}

with the VEV patterns of the scalar fields $\rho $, $\varphi $, $\Omega _{1}$%
, $\Omega _{2}$\ and $\Omega _{3}$ 
\begin{equation}
\left\langle \rho \right\rangle =v_{\rho }\left( i,0,0\right) ,\quad \quad
\left\langle \varphi \right\rangle =v_{\varphi }\left( 1,0,0\right) \,,\quad
\quad \left\langle \Omega _{1}\right\rangle =v_{\Omega _{1}},\quad \quad
\left\langle \Omega _{2}\right\rangle =v_{\Omega _{2}}e^{i\theta _{\Omega
}},\quad \quad \left\langle \Omega _{3}\right\rangle =v_{\Omega _{3}}.\,\,
\label{vsq}
\end{equation}
The $Z_{12}$ symmetry creates a hierarchy among the columns of the quark
mass matrices which leads to the quark mass hierarchies observed in
experiments without additional fine-tuning. 
The relevant $S_{4}\otimes Z_{2}^{\prime \prime }\otimes Z_{6}\otimes Z_{12}$
-invariant Yukawa terms for the up- and down-type quarks are given in Appendix %
\ref{a:quarks}. Using the $S_{4}$ multiplication rules listed in Appendix \ref%
{a:rules}, it follows that the quark mass matrices are given by 
\begin{equation}
M_{q}=\left( 
\begin{array}{ccc}
C_{q}e^{i\theta _{1q}} & 0 & 0 \\ 
D_{q}e^{-i\theta _{2q}} & E_{q}e^{-i\theta _{3q}} & F_{q}e^{-i\theta _{4q}}
\\ 
D_{q}e^{i\theta _{2q}} & E_{q}e^{i\theta _{3q}} & F_{q}e^{i\theta _{4q}}%
\end{array}%
\right) ,\quad \quad q=U,D.  \label{Mq}
\end{equation}

Then the quark mass matrices satisfy the following relation: 
\begin{eqnarray}
M_{q}M_{q}^{\dagger } &=&\left( 
\begin{array}{ccc}
C_{q}e^{i\theta _{1q}} & 0 & 0 \\ 
D_{q}e^{-i\theta _{2q}} & E_{q}e^{-i\theta _{3q}} & F_{q}e^{-i\theta _{4q}}
\\ 
D_{q}e^{i\theta _{2q}} & E_{q}e^{i\theta _{3q}} & F_{q}e^{i\theta _{4q}}%
\end{array}%
\right) \left( 
\begin{array}{ccc}
C_{q}e^{-i\theta _{1q}} & D_{q}e^{i\theta _{2q}} & D_{q}e^{-i\theta _{2q}}
\\ 
0 & E_{q}e^{i\theta _{3q}} & E_{q}e^{-i\theta _{3q}} \\ 
0 & F_{q}e^{i\theta _{4q}} & F_{q}e^{-i\theta _{4q}}%
\end{array}%
\right) \\
&=&\allowbreak \left( 
\begin{array}{ccc}
C_{q}^{2} & C_{q}D_{q}e^{i\left( \theta _{1q}+\theta _{2q}\right) } & 
C_{q}D_{q}e^{i\left( \theta _{1q}-\theta _{2q}\right) } \\ 
C_{q}D_{q}e^{-i\left( \theta _{1q}+\theta _{2q}\right) } & 
D_{q}^{2}+E_{q}^{2}+F_{q}^{2} & D_{q}^{2}e^{-2i\theta
_{2q}}+E_{q}^{2}e^{-2i\theta _{3q}}+F_{q}^{2}e^{-2i\theta _{4q}} \\ 
C_{q}D_{q}e^{i\left( \theta _{1q}-\theta _{2q}\right) } & 
D_{q}^{2}e^{2i\theta _{2q}}+E_{q}^{2}e^{2i\theta _{3q}}+F_{q}^{2}e^{2i\theta
_{4q}} & D_{q}^{2}+E_{q}^{2}+F_{q}^{2}%
\end{array}%
\right) \\
&=&\left( 
\begin{array}{ccc}
X_{q} & Y_{q}e^{i\theta _{aq}} & Y_{q}e^{i\theta _{bq}} \\ 
Y_{q}e^{-i\theta _{aq}} & U_{q} & V_{q}e^{i\theta _{cq}} \\ 
Y_{q}e^{-i\theta _{bq}} & V_{q}e^{-i\theta _{cq}} & U_{q}%
\end{array}%
\right)
\end{eqnarray}
where $X_{q}$, $Y_{q}$, $V_{q}$ and $U_{q}$ are real parameters. For the
sake of simplicity we assume $\theta _{cq}=\theta _{bq}-\theta _{aq}$, so
that the relevant physical part of the quark mass matrices can be rewritten
as follows: 
\begin{equation}
M_{q}M_{q}^{\dagger }=P_{q}J_{q}P_{q}^{\dagger },\hspace{1cm}P_{q}=\left( 
\begin{array}{ccc}
1 & 0 & 0 \\ 
0 & e^{-i\theta _{aq}} & 0 \\ 
0 & 0 & e^{-i\theta _{cq}}%
\end{array}%
\right) ,\hspace{1cm}J_{q}=\left( 
\begin{array}{ccc}
X_{q} & Y_{q} & Y_{q} \\ 
Y_{q} & U_{q} & V_{q} \\ 
Y_{q} & V_{q} & U_{q}%
\end{array}%
\right) .
\end{equation}

The matrix $J_{q}$ corresponds to a modification of the Fukuyama-Nishiura
texture proposed in \cite{Hernandez:2014zsa} 
and is diagonalized by an orthogonal matrix $R_{q}$ as follows:

\begin{equation}
R_{q}J_{q}R_{q}^{T}=diag\left(
-m_{q_{1}}^{2},m_{q_{2}}^{2},m_{q_{3}}^{2}\right) ,\hspace{1cm}\hspace{1cm}%
R_{q}=\left( 
\begin{array}{ccc}
c_{q} & s_{q} & 0 \\ 
-\frac{s_{q}}{\sqrt{2}} & \frac{c_{q}}{\sqrt{2}} & -\frac{1}{\sqrt{2}} \\ 
-\frac{s_{q}}{\sqrt{2}} & \frac{c_{q}}{\sqrt{2}} & \frac{1}{\sqrt{2}}%
\end{array}%
\right) ,
\end{equation}%
where: 
\begin{equation}
c_{q}=\sqrt{\frac{m_{q_{2}}^{2}-X_{q}}{m_{q_{2}}^{2}+m_{q_{1}}^{2}}},\hspace{%
1cm}\hspace{1cm}s_{q}=\sqrt{\frac{m_{q_{1}}^{2}+X_{q}}{%
m_{q_{2}}^{2}+m_{q_{1}}^{2}}}
\end{equation}%
with the quark masses 
\begin{eqnarray}
-m_{q_{1}}^{2} &=&\frac{1}{2}\left( U_{q}+V_{q}+X_{q}-\sqrt{\left(
X_{q}-U_{q}-V_{q}\right) ^{2}+8Y_{q}^{2}}\right) , \\
m_{q_{2}}^{2} &=&\frac{1}{2}\left( U_{q}+V_{q}+X_{q}+\sqrt{\left(
X_{q}-U_{q}-V_{q}\right) ^{2}+8Y_{q}^{2}}\right) , \\
m_{q_{3}}^{2} &=&U_{q}-V_{q}.
\end{eqnarray}%
Furthermore, for the CKM quark mixing matrix we obtain 
\begin{equation}
K=O_{U}^{T}P_{UD}O_{D}=\left( 
\begin{array}{ccc}
c_{U}c_{D}+\frac{1}{2}s_{U}s_{D}\left( e^{i\vartheta }+e^{i\varrho }\right)
& c_{U}s_{D}-\frac{1}{2}s_{U}c_{D}\left( e^{i\vartheta }+e^{i\varrho }\right)
& \frac{1}{2}s_{U}\left( e^{i\vartheta }-e^{i\varrho }\right) \\ 
s_{U}c_{D}-\frac{1}{2}c_{U}s_{D}\left( e^{i\vartheta }+e^{i\varrho }\right)
& s_{U}s_{D}+\frac{1}{2}c_{U}c_{D}\left( e^{i\vartheta }+e^{i\varrho }\right)
& \frac{1}{2}c_{U}\left( e^{i\varrho }-e^{i\vartheta }\right) \\ 
\frac{1}{2}s_{D}\left( e^{i\vartheta }-e^{i\varrho }\right) & \frac{1}{2}%
c_{D}\left( e^{i\varrho }-e^{i\vartheta }\right) & \frac{1}{2}\left(
e^{i\vartheta }+e^{i\varrho }\right)%
\end{array}%
\right) ,\allowbreak
\end{equation}%
where $P_{UD}=P_{U}^{\dagger }P_{D}=diag\left( 1,e^{i\vartheta },e^{i\varrho
}\right) $, with $\vartheta =\theta _{aU}-\theta _{aD}$ and $\varrho =\theta
_{bU}-\theta _{bD}$.

Using the values of the quark masses at the $M_{Z}$ scale shown in Table \ref%
{Quarkmasses} and\ varying the parameters $X_{U,D}$, $\vartheta $ and $%
\varrho $ we fit the magnitudes of the CKM matrix elements, the CP-violating
phase and the Jarlskog invariant $J$ to the experimental values shown in
Table \ref{Observables}. The values of the quark masses at the $M_{Z}$ scale
have been taken from Ref. \cite{Bora:2012tx}, whereas the experimental
values of the CKM magnitudes and the Jarlskog invariant $J$ are taken from
Ref. \cite{Beringer:1900zz}.

\begin{table}[tbh]
\begin{center}
\begin{tabular}{|c|l|}
\hline
Quark masses & Experimental Value \\ \hline
$m_{d}(\text{MeV})$ & $2.9_{-0.4}^{+0.5}$ \\ \hline
$m_{s}(\text{MeV})$ & $57.7_{-15.7}^{+16.8}$ \\ \hline
$m_{b}(\text{MeV})$ & $2820_{-40}^{+90}$ \\ \hline
$m_{u}(\text{MeV})$ & $1.45_{-0.45}^{+0.56}$ \\ \hline
$m_{c}(\text{MeV})$ & $635\pm 86$ \\ \hline
$m_{t}(\text{MeV})$ & $172.1\pm 0.6\pm 0.9$ \\ \hline
\end{tabular}%
\end{center}
\caption{Experimental values of the quark masses at the $M_{Z}$ scale. 
\protect\cite{Bora:2012tx}.}
\label{Quarkmasses}
\end{table}

\begin{table}[tbh]
\begin{center}
\begin{tabular}{|c|l|l|}
\hline
Observable & Fukuyama like texture & Experimental Value \\ \hline
$\bigl|V_{ud}\bigr|$ & \quad $0.974$ & \quad $0.97427\pm 0.00015$ \\ \hline
$\bigl|V_{us}\bigr|$ & \quad $0.225$ & \quad $0.22534\pm 0.00065$ \\ \hline
$\bigl|V_{ub}\bigr|$ & \quad $0.00351$ & \quad $%
0.00351_{-0.00014}^{+0.00015} $ \\ \hline
$\bigl|V_{cd}\bigr|$ & \quad $0.225$ & \quad $0.22520\pm 0.00065$ \\ \hline
$\bigl|V_{cs}\bigr|$ & \quad $0.973$ & \quad $0.97344\pm 0.00016$ \\ \hline
$\bigl|V_{cb}\bigr|$ & \quad $0.0412$ & \quad $0.0412_{-0.0005}^{+0.0011}$
\\ \hline
$\bigl|V_{td}\bigr|$ & \quad $0.00867$ & \quad $%
0.00867_{-0.00031}^{+0.00029} $ \\ \hline
$\bigl|V_{ts}\bigr|$ & \quad $0.0404$ & \quad $0.0404_{-0.0005}^{+0.0011}$
\\ \hline
$\bigl|V_{tb}\bigr|$ & \quad $0.999$ & \quad $%
0.999146_{-0.000046}^{+0.000021}$ \\ \hline
$J$ & \quad $2.96\times 10^{-5}$ & \quad $(2.96_{-0.16}^{+0.20})\times
10^{-5}$ \\ \hline
$\delta $ & \quad $69.2^{\circ }$ & \quad $68^{\circ }$ \\ \hline
\end{tabular}%
\end{center}
\caption{Experimental CKM magnitudes and Jarlskog invariant compared with
results obtained from our fit.}
\label{Observables}
\end{table}

Using the values 
\begin{equation}
X_{U}=2.90\times 10^{-3}\,\text{GeV}^{2},\hspace{1cm}X_{D}=1.38\times
10^{-4}\,\text{GeV}^{2},\hspace{1cm}\vartheta =87.9^{\circ },\hspace{1cm}%
\varrho =92.6^{\circ },
\end{equation}%
the obtained magnitudes of the CKM matrix elements, the CP-violating phase
and the Jarlskog invariant are in excellent agreement with the experimental
data.
\vspace{1cm}

\section{$Z_3$ breaking}

\label{s:perturb}

\subsection{Scalar sector}

The $S_4$ symmetry of the model is broken down to a residual $Z_3$ symmetry
once the $S_4$ triplet $\phi$ acquires VEVs in the direction $v(1,1,1)$.
However, perturbations can arise from other scalars without a mechanism to
protect the necessary VEV alignments, e.g., from a scalar triplet $\chi$
acquiring VEVs in the $v_{\chi}(1,0,0)$ direction to generate neutrino masses and mixings (%
cf. Eqs. (\ref{e:numasses}--\ref{e:vevsnu})). These perturbations
are caused by quartic interactions of the form $(\phi^\dagger
\phi)(\chi^\dagger \chi)$ that appear in the scalar potential and cannot be
forbidden by a flavor symmetry since the combinations $\phi^\dagger \phi$
and $\chi^\dagger \chi$ are always invariant.

In our model the scalar triplet responsible for the charged lepton masses is 
$\phi $ with $\langle \phi \rangle =v(1,1,1)$, assigned to $(3^{\prime },2)$
under $S_{4}\otimes SU(2)$, whereas several scalars responsible for mixings
in the quark and neutrino sector cause deviations from this alignment
through quartic interactions with $\phi $. If we assume a VEV hierarchy
among those scalars to simplify the discussion, i.e., $v_{\rho },v_{\varphi }\gg v_{\chi },v_{\xi
},v_{\sigma },v_{\zeta }$, the perturbations coming from scalars involved in
neutrino interactions can be neglected. The remaining fields being 
\begin{equation*}
\rho :~(3^{\prime },1)\,,\qquad \varphi ~:~(3,1)\,\qquad \text{with}\qquad
\langle \rho \rangle =v_{\rho }\left( i,0,0\right) \,,\qquad \left\langle
\varphi \right\rangle =v_{\varphi }\left( 1,0,0\right) .
\end{equation*}%
Thus, the relevant cross couplings in the scalar potential are 
\begin{equation}
V_{\text{int}}\supset \sum_{i=\mathbf{1},\mathbf{2},\mathbf{3},\mathbf{%
3^{\prime }}}(\phi ^{\dagger }\phi )_{i}\left[ \lambda _{\rho _{i}}(\rho
^{\dagger }\rho )_{i}+\lambda _{\varphi _{i}}(\varphi ^{\dagger }\varphi
)_{i}\right] ,  \label{e:vint}
\end{equation}

where $i=\mathbf{1},\mathbf{2},\mathbf{3},\mathbf{3^\prime}$ denotes the
corresponding $S_4$ contraction. Eventually only the $\mathbf{2}$%
-contractions, e.g., $\sum^3_{j,k=1, j\neq k} 2|\phi_j|^2 |\rho_j|^2 -
|\phi_j|^2 |\rho_k|^2$, result in perturbations of the $Z_3$ conserving VEV
alignment $v(1,1,1)$ as the other contractions are invariant under the $Z_3$
conserving generator after the scalars $\rho$ and $\varphi$ acquire their
VEVs. Therefore we only need to consider the following terms in the scalar
potential to analyze the breaking of $Z_3$ 
\begin{align}
V_{\text{int}} &\supset (\phi^\dagger \phi)_{\mathbf{2}} \left[
\lambda_{\rho} (\rho^\dagger \rho)_{\mathbf{2}} + \lambda_{\varphi}
(\varphi^\dagger \varphi)_{\mathbf{2}} \right] \\
&= \sum^3_{j,k=1, j\neq k} |\phi_j|^2 \left[\lambda_{\rho}(2 |\rho_j|^2 -
|\rho_k|^2) + \lambda_{\varphi}(2 |\varphi_j|^2 - |\varphi_k|^2) \right].
\end{align}
Assuming for simplicity that the coupling constants are the same order of
magnitude, i.e., $\lambda_{\rho} \approx \lambda_{\varphi} \approx
\lambda_s/2$, the VEV alignment of the triplet $\phi$ is approximately
shifted by a perturbation $\epsilon$ in the following way 
\begin{align}
\langle\phi\rangle = v(1+2\epsilon,1-\epsilon,1-\epsilon),
\end{align}
where the contributions from $\rho$ and $\varphi$ are summarized in the
parameter $\epsilon$. 
Doing so, we adopt a similar approach as in \cite{Heeck:2014qea}, who recently analyzed
a triality model based on an $A_4$ flavor symmetry.

As a consequence, one of the physical Higgs doublets which was initially
inert before the breaking of $Z_3$, $\langle \phi_b \rangle = 0$, now
acquires a small VEV depending on the size of the perturbation parameter $%
\epsilon$ 
\begin{align}
\left( 
\begin{array}{c}
\langle \phi_a \rangle \\ 
\langle \phi_b \rangle \\ 
\langle \phi_c \rangle%
\end{array}
\right) = \left( 
\begin{array}{ccc}
0 & -\frac{1}{\sqrt{2}} & \frac{1}{\sqrt{2}} \\ 
\frac{\sqrt{2}}{\sqrt{3}} & -\frac{1}{\sqrt{6}} & -\frac{1}{\sqrt{6}} \\ 
\frac{1}{\sqrt{3}} & \frac{1}{\sqrt{3}} & \frac{1}{\sqrt{3}}%
\end{array}
\right) \left( 
\begin{array}{c}
\frac{v}{\sqrt{3}}(1+2\epsilon) \\ 
\frac{v}{\sqrt{3}}(1-\epsilon) \\ 
\frac{v}{\sqrt{3}}(1-\epsilon)%
\end{array}
\right) = v \left( 
\begin{array}{c}
0 \\ 
\sqrt{2} \epsilon \\ 
1%
\end{array}
\right).
\end{align}
Following \cite{Heeck:2014qea} we use the parametrization 
\begin{align}
\langle (\phi_a, \phi_b, \phi_c)^T \rangle = v(0, \sin \theta, \cos \theta),
\end{align}
where $\theta$ is given by a combination of parameters from the scalar
potential to account for the deviation from LFT. From now onwards we will
use the abbreviations $\sin \theta := s_\theta$ and $\cos \theta := c_\theta$%
.

The breaking of $Z_3$ induces new mixing of the doublets $\phi_b$ and $%
\phi_c $ which were initially mass eigenstates. The CP-odd neutral scalars $%
\phi_{(b,c),I}^0$ and the charged scalars $\phi_{b,c}^+$ mix via 
\begin{align}
\left( 
\begin{array}{c}
H^\pm \\ 
\pi^\pm%
\end{array}
\right) = \left( 
\begin{array}{cc}
c_{\theta} & s_{\theta} \\ 
- s_{\theta} & c_{\theta}%
\end{array}
\right) \left( 
\begin{array}{c}
\phi_b^\pm \\ 
\phi_c^\pm%
\end{array}
\right) \quad \text{and} \quad \left( 
\begin{array}{c}
\eta^0_I \\ 
\pi^0_I%
\end{array}
\right) = \left( 
\begin{array}{cc}
c_{\theta} & s_{\theta} \\ 
- s_{\theta} & c_{\theta}%
\end{array}
\right) \left( 
\begin{array}{c}
\phi_{b,I}^0 \\ 
\phi_{c,I}^0%
\end{array}
\right),
\end{align}
where $\pi^\pm$ and $\pi^0_I$ are massless goldstone bosons. In the case of
the CP-even neutral scalars the situation is more complicated and the mixing
results in a mass splitting of the scalars which were initially degenerate in mass. The complete mass spectrum reads 
\begin{align}
m^2_{\phi^0_{a,I}} &= m^2_{\eta^0_I} = -v^2 \delta, \qquad m^2_{\pi^0_I} = 0,
\\
m^2_{\phi^{\pm}_{a}} &= m^2_{H^{\pm}} = -v^2(\kappa - \beta), \qquad
m^2_{\pi^\pm} = 0, \\
m^2_{\phi^{0}_{a,R}} &= \frac{1}{6} v^2 (-2 + 2 \sqrt{2} c_{\theta}
s_{\theta} + s_{\theta}^2) \kappa , \\
m^2_{h} &= \frac{1}{2} (v^2 \alpha - m^2_{\phi^{0}_{a,R}}) + \Delta, \qquad
m^2_{H} = \frac{1}{2} (v^2 \alpha - m^2_{\phi^{0}_{a,R}}) - \Delta \\
\text{with} \qquad \Delta &= \frac{1}{6} v^2 \sqrt{9(\alpha + \kappa)^2+ 3
\kappa s_{\theta} \left[2 \sqrt{2} c_{\theta} (\alpha + \kappa) - 3
s_{\theta} (5\alpha + 3 \kappa) \right] + \mathcal{O}(s^3_{\theta})}.
\end{align}
Note that only the masses of the CP-even scalars depend on the perturbation
parameter $\theta$. Hence $Z_3$ breaking in this direction does not affect
the phenomenology of the CP-odd and charged scalars in our triality model.
In the triality limit $\theta \rightarrow 0$ the mixing vanishes and the
original mass spectrum is recovered with $m^2_{h} 
\xrightarrow{\theta
\rightarrow 0} m^2_{\phi_{c,R}^0}$ and $m^2_{H} 
\xrightarrow{\theta
\rightarrow 0} m^2_{\phi_{b,R}^0} = m^2_{\phi_{a,R}^0}$. Therefore the
scalar $h$ should play the role of the SM-like Higgs with $m_h \approx 125\,$%
GeV, whereas $H$ will be a new heavy Higgs as in regular
two-Higgs-doublet models (2HDM). The mixing angle $\vartheta$ between the CP
even neutral scalars is defined by 
\begin{align}
\left( 
\begin{array}{c}
H \\ 
h%
\end{array}
\right) &= \left( 
\begin{array}{cc}
c_{\vartheta} & s_{\vartheta} \\ 
- s_{\vartheta} & c_{\vartheta}%
\end{array}
\right) \left( 
\begin{array}{c}
\phi_{b,R}^0 \\ 
\phi_{c,R}^0%
\end{array}
\right) \qquad \text{and} \\
\tan 2\vartheta &= -\frac{4 s_\theta (\sqrt{2} \kappa s_\theta - 2 c_\theta
(3 \alpha + \kappa))}{ 6 (-1 + 2 s_\theta^2) \alpha + (-6 - 2 \sqrt{2}
c_\theta s_\theta + 11 s_\theta^2) \kappa}
\end{align}
with $\vartheta \rightarrow 0$ for $\theta \rightarrow 0$ in the unbroken
triality limit. In terms of the scalar masses $m_h \approx 125\,$GeV and $%
m_{\phi^{0}_{a,R}} := m_a$ this is 
\begin{align}  \label{e:tan2vt}
\tan 2\vartheta &= \frac{A}{B} \qquad \text{with} \\
A &= s_\theta (c_\theta (4 m_a^2 m_h^2 s_\theta^2 (2 - 7 s_\theta^2) + 4
m_h^4 (4 + 4 s_\theta^2 - 7 s_\theta^4) - 8 m_a^4 (2 - 5 s_\theta^2 + 3
s_\theta^4))  \notag \\
&- 4 \sqrt{2} s_\theta (4 m_h^4 (2 - 3 s_\theta^2 + s_\theta^4) + m_a^2
m_h^2 (2 - 7 s_\theta^2 + 4 s_\theta^4) + m_a^4 (6 - 19 s_\theta^2 + 15
s_\theta^4))),  \notag \\
B &= 2 m_a^2 m_h^2 (4 - 16 s_\theta^2 + 17 s_\theta^4 + 4 \sqrt{2} c_\theta
s_\theta^5 - 7 s_\theta^6)  \notag \\
&+ m_h^4 (-1 + 2 s_\theta^2) (4 + 4 s_\theta^2 - 7 s_\theta^4 + 4 \sqrt{2}
c_\theta s_\theta (-2 + s_\theta^2))  \notag \\
&+ m_a^4 (-4 + 12 s_\theta^2 - 11 s_\theta^4 + 5 s_\theta^6 - 2 \sqrt{2}
c_\theta s_\theta (4 - 14 s_\theta^2 + 13 s_\theta^4)).  \notag
\end{align}
Since the mass of the scalar $h$ is fixed, the mixing angle $\vartheta$
depends only on two parameters. Here we chose the perturbation angle $\theta$
and the mass of the exotic Higgs $\phi_{a,R}^0$. Through the relation $m_a^2
= v^2 \alpha - (m_h^2 + m_H^2)$ one can alternatively display the results in
terms of $m_H$, which is the heavy Higgs in our pseudo 2HDM.

\subsection{Consequences for lepton mixing}

As a consequence of the perturbed alignment $v(0,s_{\theta}, c_{\theta})$
the mass matrix of the charged leptons takes the form 
\begin{align}
M_l&=\frac{v}{\sqrt{6}}\left( 
\begin{array}{ccc}
\left(\text{$c_\theta $}+\sqrt{2} \text{$s_\theta $}\right) (\widetilde{y}_2-%
\widetilde{y}_3)\lambda^5 & \left(\text{$c_\theta $}+\sqrt{2} \text{$%
s_\theta $}\right) (\widetilde{y}_2+\widetilde{y}_3)\lambda^5 & \left(\text{$%
c_\theta $}+\sqrt{2} \text{$s_\theta $}\right) \widetilde{y}_1 \lambda^3 \\ 
\frac{1}{2} \left(2 \text{$c_\theta $}-\sqrt{2} \text{$s_\theta $}\right) (%
\widetilde{y}_2-\widetilde{y}_3)\lambda^5 \omega^2 & \frac{1}{2} \left(2 
\text{$c_\theta $}-\sqrt{2} \text{$s_\theta $}\right) (\widetilde{y}_2+%
\widetilde{y}_3)\lambda^5 \omega & \frac{1}{2} \left(2 \text{$c_\theta $}-%
\sqrt{2} \text{$s_\theta $}\right) \widetilde{y}_1 \lambda^3 \\ 
\frac{1}{2} \left(2 \text{$c_\theta $}-\sqrt{2} \text{$s_\theta $}\right) (%
\widetilde{y}_2-\widetilde{y}_3)\lambda^5 \omega & \frac{1}{2} \left(2 \text{%
$c_\theta $}-\sqrt{2} \text{$s_\theta $}\right) (\widetilde{y}_2+\widetilde{y%
}_3)\lambda^5 \omega^2 & \frac{1}{2} \left(2 \text{$c_\theta $}-\sqrt{2} 
\text{$s_\theta $}\right) \widetilde{y}_1 \lambda^3 \\ 
&  & 
\end{array}
\right) \\
&= \frac{v}{\sqrt{2}} \underbrace{\frac{1}{\sqrt{3}}\left( 
\begin{array}{ccc}
1 & 1 & 1 \\ 
\omega^2 & \omega & 1 \\ 
\omega & \omega^2 & 1%
\end{array}
\right)}_{U_{\omega}} \left( 
\begin{array}{ccc}
\text{$c_\theta $} (\widetilde{y}_2-\widetilde{y}_3)\lambda^5 & \frac{\text{$%
s_\theta $} (\widetilde{y}_2+\widetilde{y}_3)\lambda^5}{\sqrt{2}} & \frac{%
\text{$s_\theta $}\widetilde{y}_1 \lambda^3 }{\sqrt{2}} \\ 
\frac{\text{$s_\theta $} (\widetilde{y}_2-\widetilde{y}_3)\lambda^5}{\sqrt{2}%
} & \text{$c_\theta $} (\widetilde{y}_2+\widetilde{y}_3)\lambda^5 & \frac{%
\text{$s_\theta $} \widetilde{y}_1 \lambda^3 }{\sqrt{2}} \\ 
\frac{\text{$s_\theta $} (\widetilde{y}_2-\widetilde{y}_3)\lambda^5}{\sqrt{2}%
} & \frac{\text{$s_\theta $} (\widetilde{y}_2+\widetilde{y}_3)\lambda^5}{%
\sqrt{2}} & \text{$c_\theta $} \widetilde{y}_1 \lambda^3%
\end{array}
\right),
\end{align}
hence with off-diagonal elements that vanish in the limit $\theta
\rightarrow 0$ it is no longer diagonal in the $Z_3$ basis of the charged
leptons. The perturbed charged lepton mass matrix can be diagonalized to a
good approximation by 
\begin{align}
V_L^{\dagger} M_l V_R = \text{diag}(m_e, m_{\mu}, m_{\tau})
\end{align}
with 
\begin{equation}
\begin{aligned} \label{e:vrl} V_R &\simeq \left( \begin{array}{ccc} 1 &
-\frac{\sqrt{2} \text{$s_\theta $}
(\widetilde{y}_2-\widetilde{y}_3)}{(\widetilde{y}_2+\widetilde{y}_3)} &
-\frac{\sqrt{2} \text{$s_\theta $}
(\widetilde{y}_2-\widetilde{y}_3)\lambda^2}{\widetilde{y}_1 } \\
\frac{\sqrt{2} \text{$s_\theta $}
(\widetilde{y}_2-\widetilde{y}_3)}{(\widetilde{y}_2+\widetilde{y}_3)} & 1 &
-\frac{\sqrt{2} \text{$s_\theta $}
(\widetilde{y}_2+\widetilde{y}_3)\lambda^2}{\widetilde{y}_1 } \\
\frac{\sqrt{2} \text{$s_\theta $}
(\widetilde{y}_2-\widetilde{y}_3)\lambda^2}{\widetilde{y}_1 } &
\frac{\sqrt{2} \text{$s_\theta $}
(\widetilde{y}_2+\widetilde{y}_3)\lambda^2}{\widetilde{y}_1 } & 1
\end{array} \right), \\ V_L &\simeq U_{\omega} R O_{23}(\theta) R^T
O_{12}(\alpha_L), \qquad R = \left( \begin{array}{ccc} -\frac{1}{\sqrt{2}} &
\frac{1}{\sqrt{2}} & 0 \\ \frac{1}{\sqrt{2}} & \frac{1}{\sqrt{2}} & 0 \\ 0 &
0 & 1 \end{array} \right), \end{aligned}
\end{equation}

where $O_{ij}$ are rotation matrices in the $ij$-plane and 
\begin{align}  \label{e:tan2al}
\tan 2 \alpha_L=\frac{\text{$s_\theta $} \left(4 \sqrt{2} \text{$c_{3\theta} 
$}+12 \sqrt{2} \text{$c_\theta $}-7 \text{$s_{3\theta} $}-3 \text{$s_\theta $%
}\right)}{10 \text{$c_{3\theta} $}+6 \text{$c_\theta $}+8 \sqrt{2} \text{$%
s_\theta $}^3}.
\end{align}

Using the rotation matrices (Eq. \ref{e:vrl})) we find the charged lepton
Yukawa couplings $y_{ll^{\prime}} (l,l^{\prime} = e,\mu,\tau)$ that
eventually lead to flavor violation in the lepton sector. With 
\begin{align}
(\overline{L}_{1},\overline{L}_{2},\overline{L}_{3}) = (\overline{L}_{e},%
\overline{L}_{\mu},\overline{L}_{\tau}) V_L^{\dagger}, \qquad (e_1, e_2,
e_3)^T = V_R \cdot (e,\mu,\tau)^T, \qquad (\phi_1, \phi_2, \phi_3)^T = U_s
\cdot (\phi_a, H, h)^T,
\end{align}
\begin{align}
\text{where} \quad U_s \equiv \left( 
\begin{array}{ccc}
0 & \sqrt{\frac{2}{3}} & \frac{1}{\sqrt{3}} \\ 
-\frac{1}{\sqrt{2}} & -\frac{1}{\sqrt{6}} & \frac{1}{\sqrt{3}} \\ 
\frac{1}{\sqrt{2}} & -\frac{1}{\sqrt{6}} & \frac{1}{\sqrt{3}}%
\end{array}
\right) \cdot \left( 
\begin{array}{ccc}
1 & 0 & 0 \\ 
0 & \text{$c_\vartheta $} & -\text{$s_\vartheta $} \\ 
0 & \text{$s_\vartheta $} & \text{$c_\vartheta $}%
\end{array}
\right),
\end{align}
we can identify the coefficients of $h l_L l^{\prime}_R$ as the Yukawas $%
y_{l l^{\prime}}$. In leading order, i.e., $m_e \ll m_{\mu} \ll m_{\tau}$,
the dominant flavor violating couplings are 
\begin{align}  \label{e:ymt}
y_{e\tau} &\simeq -\frac{m_{\tau}}{v} (c_{\alpha_L} - s_{\alpha_L})
s_{\vartheta + \theta}, \\
y_{\mu\tau} &\simeq -\frac{m_{\tau}}{v} (c_{\alpha_L} + s_{\alpha_L})
s_{\vartheta + \theta}, \\
y_{e\mu} &\simeq \frac{m_{\mu}}{4v} \left( 2 \text{$c_\vartheta $} \left(\text{$%
c_{\alpha_L} $} \left(\text{$c_\theta $}-2 \text{$s_\theta $}^2-1\right)-%
\text{$s_{\alpha_L} $} \left(\text{$c_\theta $}-2 \text{$s_\theta $}%
^2+1\right)\right) \right. \\
&\qquad~- ~\text{$s_\vartheta $} \left. \left(\text{$c_{\alpha_L} $} \left(%
\text{$c_\theta $} \left(4 \text{$s_\theta $}+\sqrt{2}\right)-2 \text{$%
s_\theta $}+\sqrt{2}\right)+\text{$s_{\alpha_L} $} \left(-\text{$c_\theta $}
\left(4 \text{$s_\theta $}+\sqrt{2}\right)+2 \text{$s_\theta $}+\sqrt{2}%
\right)\right) \right).  \notag
\end{align}
The remaining off-diagonal couplings are negligibly small and therefore
irrelevant for the discussion 
\begin{align}
y_{\tau e} \ll y_{e\tau}, \qquad y_{\mu e} \ll y_{e \mu}, \qquad y_{\tau
\mu} \ll y_{\mu \tau},
\end{align}
which means that $y_{\mu\tau}$ dominates the $h \rightarrow \mu \tau$ decay
channel in our model. 
The perturbation of the VEV alignment has an effect on the PMNS mixing
matrix that is negligible for sufficiently small values of $\theta $. This
extra contribution modifies the charged lepton contribution to the PMNS
mixing matrix as follows 
\begin{equation}
V_{L}=U_{\omega }W_{L}\quad \text{with}\quad W_{L}\simeq \left( 
\begin{array}{ccc}
1 & \frac{\theta }{\sqrt{2}}-\frac{3\theta ^{2}}{4} & \frac{\theta }{\sqrt{2}%
} \\ 
-\frac{\theta }{\sqrt{2}}+\frac{3\theta ^{2}}{4} & 1 & \frac{\theta }{\sqrt{2%
}} \\ 
-\frac{\theta }{\sqrt{2}} & -\frac{\theta }{\sqrt{2}} & 1%
\end{array}%
\right) .
\end{equation}%

As we will explain in Sec. \ref{s:lfvhiggs} the values required to explain
the $h\rightarrow \mu \tau $ measurement are in the region $|\theta|
\lesssim 0.10$ @ 95\%~\text{C.L.} The leptonic mixing angles on the other hand stay within
their experimentally determined $3\,\sigma $ ranges for $\theta <0.08$, but
deviations in $\theta _{12}$ quickly become too large for increasing values
of $\theta $. The full PMNS matrix and the deviations in the mixing angles caused by the perturbation $\theta$ are given in Eqs. (\ref{PMNSZ3B}) and (\ref{e:sin12p}) - (\ref{e:fth}) of Appendix \ref{a:mix}.

The breaking of $Z_{3}$ also affects the Jarlskog invariant as follows 
\begin{equation}
J\approx \frac{\cos (2\psi )}{6\sqrt{3}}+\frac{1}{12}\theta \left( \sqrt{6}%
\sin (2\psi )\cos (\phi )-\sqrt{2}\sin (2\psi )\sin (\phi )\right) +\frac{1}{%
24}\theta ^{2}\left( -3\sqrt{3}\cos (2\psi )-\sin (2\psi )\sin (\phi )+\sqrt{%
3}\right)  \label{e:jp}
\end{equation}
and therefore induces CP violation in neutrino oscillations.

By varying $\phi $, for $\theta =0.07$, we obtain the following best-fit
result: 
\begin{eqnarray}
&&\mbox{NH}\ :\ \phi =-0.446\pi ,\ \ \ \sin ^{2}\theta _{12}\approx 0.375,\
\ \ \sin ^{2}\theta _{23}\approx 0.586,\ \ \ \sin ^{2}\theta _{13}\approx
0.0232,\ \ \ J\approx 1.11\times 10^{-2}, \\[0.12in]
&&\mbox{IH}\hspace{2.5mm}:\ \phi =\ \ 0.553\,\pi ,\ \ \ \ \ \sin ^{2}\theta
_{12}\approx 0.375,\ \ \ \sin ^{2}\theta _{23}\approx 0.588,\ \ \ \ \,\sin
^{2}\theta _{13}\approx 0.0238,\ \ \ J\approx 1.10\times 10^{-2}.
\end{eqnarray}
Since the obtained Jarlskog invariant for both NH and IH is approximately $10^{-2}$, the $Z_{3}$
symmetry breaking can generate a sizeable strength of CP violation in neutrino
oscillations for larger $\theta$ values.

\subsection{Flavor-violating Higgs decays}

\label{s:lfvhiggs}

\begin{figure}[h]
\centering
\subfigure{\includegraphics[width=0.35\textwidth]{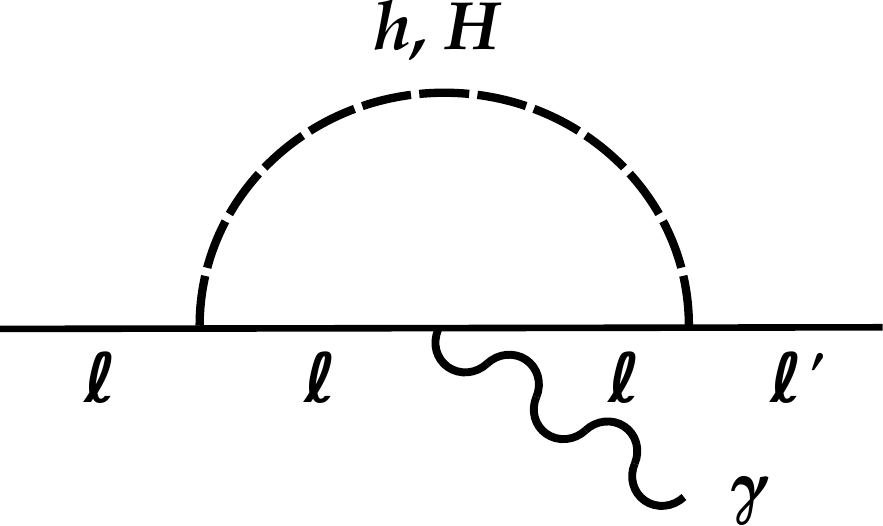}} \hspace{1cm%
} \subfigure{\includegraphics[width=0.35\textwidth]{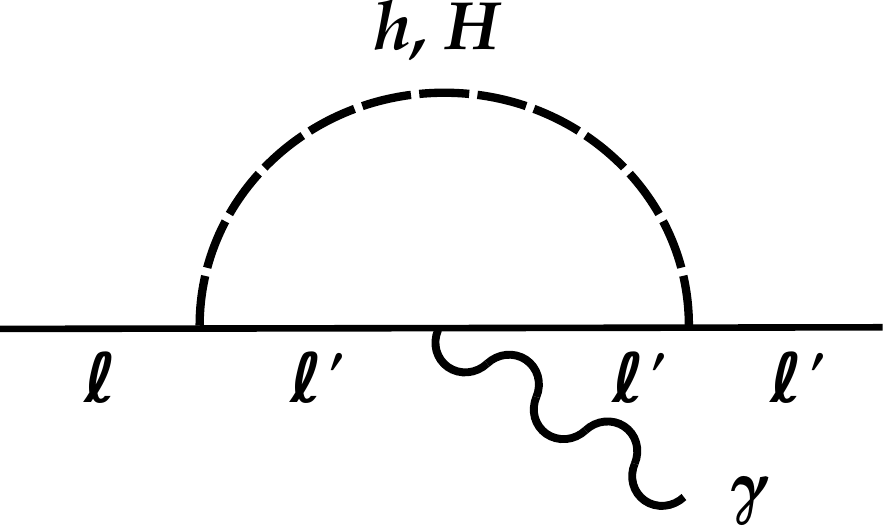}}
\caption{One-Loop diagrams contributing to $l \rightarrow l^{\prime} \protect%
\gamma$. Since $|y_{ll}| \gg |y_{l^{\prime}l^{\prime}}|$, the left diagram
is the dominating one.}
\label{fig:diags}
\end{figure}

As stated in Sec \ref{s:intro}, the measurement of $h \rightarrow \mu \tau$
is equivalent to a bound on the off-diagonal Yukawa couplings \cite%
{Dorsner:2015mja} 
\begin{align}
0.0019 (0.0008) < \sqrt{|y_{\mu\tau}|^2+|y_{\tau\mu}|^2} < 0.0032
(0.0036)~@~68\% (95\%)~\text{C.L.},  \label{e:ybound}
\end{align}
where it is assumed that $h\rightarrow \mu\tau$ is the only additional
contribution to the SM Higgs decay width. The result in Eq. (\ref{e:ybound})
is well compatible with the current bounds $\text{Br}(\tau \rightarrow \mu
\gamma) < 4.4 \times 10^{-8}~@~90\%$ C.L. and $\text{Br}(\tau \rightarrow e
\gamma) < 3.3 \times 10^{-8}~@~90\%$ \cite{PDG-2014}. However, while small
compared to $y_{\mu\tau}$, $y_{e\mu}$ dominates $h\rightarrow e\mu$ and is
tightly constrained by $\text{Br}(\mu \rightarrow e \gamma)<5.7\times
10^{-13}$ \cite{Adam:2013mnn}, which consequently restricts the allowed values of $\theta$
and $m_a$ or $m_H$. The corresponding diagrams mediating these
processes in our model are shown in Fig. \ref{fig:diags}. Analytically the
branching ratio of $l \rightarrow l^{\prime} \gamma$ is given by \cite%
{Harnik:2012pb} 
\begin{align}  \label{e:brmeg}
\text{Br}(l \rightarrow l^{\prime} \gamma) = \frac{\tau_{l} \alpha_{\text{EM}%
} m_l^5}{64 \pi^4} \left(|c_L|^2 + |c_R|^2\right),
\end{align}
where $c_L$ and $c_R$ are the Wilson coefficients. To calculate $c_L$ and $%
c_R$ one-loop (cf. Fig. \ref{fig:diags}) and two-loop contributions are
taken into account, where the latter can dominate the branching ratio in
certain regions of the parameter space. The full set of equations is shown
in Eqs. (\ref{e:cr})-(\ref{e:crw}) of Appendix \ref{a:llg} \cite%
{Harnik:2012pb,Chang:1993kw}.

\begin{figure}[h]
\centering
\subfigure[]{\includegraphics[width=0.47\textwidth]{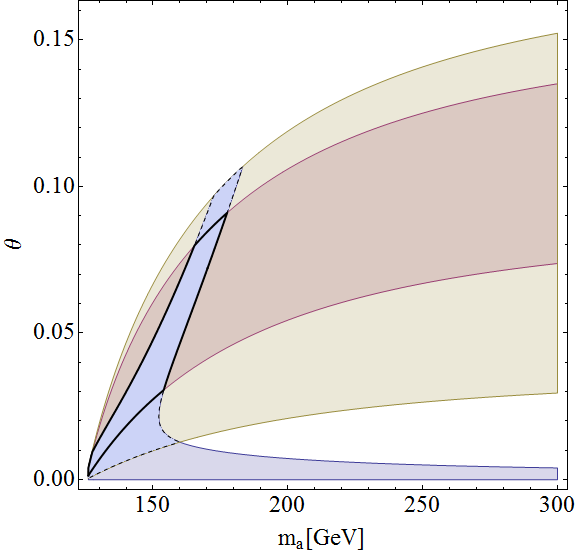}} %
\subfigure[]{\includegraphics[width=0.485\textwidth]{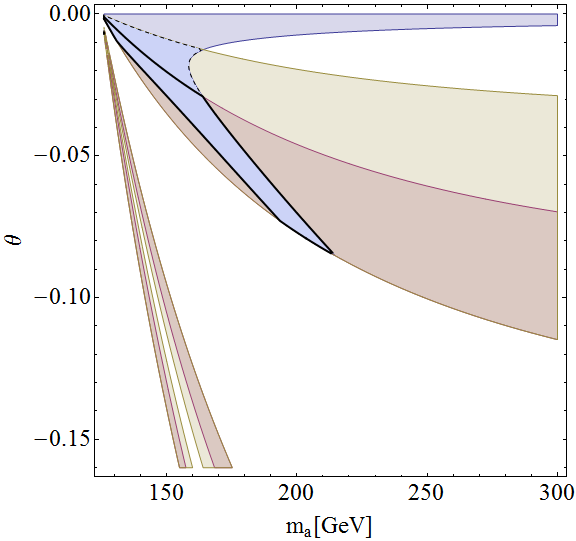}}
\caption{Parameter values leading to $|y_{\protect\mu\protect\tau}|$
required by CMS data on $h\rightarrow \protect\mu\protect\tau$ in brown
(68\%~\text{C.L.}) and yellow (95\%~\text{C.L.}) for (a) positive and (b)
negative $\protect\theta$. Parameter regions shaded in blue are allowed by $%
\text{Br}(\protect\mu \rightarrow e \protect\gamma) < 5.7 \times 10^{-13}$.
Solid (dashed) lines mark the $1\,\protect\sigma$ ($3\,\protect\sigma$)
intervals of $|y_{\protect\mu\protect\tau}|$, where both regions overlap.
Overlap occurs for $0.01 \lesssim \protect\theta \lesssim 0.09$, $126\,\text{%
GeV} \lesssim m_H \lesssim 204\,\text{GeV}$ (68\%~\text{C.L.}) and $-0.08
\lesssim \protect\theta \lesssim -0.01$, $127\,\text{GeV} \lesssim m_H
\lesssim 190\,\text{GeV}$ (68\%~\text{C.L.}) allowing us to explain the
excess in $h \rightarrow \protect\mu\protect\tau$ and at the same time to
comply with the tight bound on $\text{Br}(\protect\mu \rightarrow e \protect%
\gamma)$ ($m_{\protect\eta} = 200\,$GeV).}
\label{fig:theta}
\end{figure}

Using Eqs. (\ref{e:tan2vt}), (\ref{e:tan2al}) and (\ref{e:ymt}) we determine
numerically the allowed values of $\theta$ and $m_H$ that can explain the 2.4%
$\,\sigma$ anomaly in $h\rightarrow \mu\tau$ and at the same time respect
the bound on $\text{Br}(\mu \rightarrow e \gamma)$, which constraints the $%
y_{e\mu}$ coupling. Scanning the parameter space for negative and positive
values of $\theta$, we find that a window opens up for rather light masses
of the extra scalars $H$ and $\eta$ in the vicinity of the SM Higgs mass,
leading to overlap of the regions complying with either $\text{Br}(\mu
\rightarrow e \gamma)$ or $\text{Br}(h\rightarrow \mu\tau)$, as shown in
Figs. \ref{fig:theta}a) and \ref{fig:theta}b). This is where the
contributions of the Higgs $h$ to $\text{Br}(\mu \rightarrow e \gamma)$ are
partly canceled by the contributions of the neutral scalars $H$ and $\eta$,
allowing for larger flavor violating Yukawa couplings in these regions. The
parameter space is practically symmetric around $\theta=0$, i.e., at $68\%~%
\text{C.L.} (95\%~\text{C.L.})$ and $m_{\eta} = 200\,$GeV 
\begin{equation}  \label{e:num}
\begin{array}{cccrcl}
0.002\,(0.001) & < \theta < & 0.090\,(0.104)\,, \qquad & 126\,(126)\,\text{%
GeV} & < m_H < & 204\,(214)\,\text{GeV}, \\ 
-0.004\,(-0.001) & > \theta > & -0.082\,(-0.082)\,, \qquad & 127\,(126)\,%
\text{GeV} & < m_H < & 190\,(190)\,\text{GeV}.%
\end{array}%
\end{equation}

Using the determined $1\,\sigma$ parameter ranges (Eq. (\ref{e:num})), we can
also make other predictions for rare decays, where by far the strongest
constraints come from the radiative decays $l \rightarrow l^{\prime} \gamma$%
. The flavor violating modes $\text{Br}(\tau \rightarrow 3 \mu) < 2.1 \times
10^{-8}$, $\text{Br}(\tau \rightarrow 3 e) < 2.7 \times 10^{-8}$ and $\text{%
Br}(\mu \rightarrow 3 e) < 1.0 \times 10^{-12}$ \cite{PDG-2014} are
approximately 
\begin{align}
\text{Br}(l \rightarrow 3 l^{\prime}) = -\frac{\tau_{l} \alpha^2_{\text{EM}}
m_l^5}{72 (2\pi)^5} \left[12 \log \frac{{m_l^{\prime}}^2}{m_l^2} + 29 + 6
\log 4 \right] \left(|c_L|^2 + |c_R|^2\right),
\end{align}
assuming that loop diagrams in the spirit of Fig. \ref{fig:diags} ($\gamma$
decays into $l^+ l^-$) dominate over the tree level exchange of a neutral
scalar field \cite{Harnik:2012pb}.

Our predictions for $0.01\lesssim \theta \lesssim 0.09$, $126\,\text{GeV}%
\lesssim m_{H}\lesssim 204\,\text{GeV}$ ($m_{\eta }=200\,$GeV), 
\begin{equation}
\begin{array}{llll}
\text{Br}(\tau \rightarrow \mu \gamma ) & \in (0.3-1.3)\times 10^{-8}, & 
\qquad \text{Br}(\tau \rightarrow e\gamma ) & \in (0.3-1.3)\times 10^{-8},
\\ 
\text{Br}(\tau \rightarrow 3\mu ) & \in (1.9-8.1)\times 10^{-10}, & \qquad 
\text{Br}(\tau \rightarrow 3e) & \in (0.9-4.1)\times 10^{-9}, \\ 
\text{Br}(h\rightarrow e\tau ) & \in (0.4-1.2)\times 10^{-2}, & \qquad \text{%
Br}(h\rightarrow e\mu ) & \in (1.5-4.2)\times 10^{-5},%
\end{array}%
\end{equation}%
and $-0.08\lesssim \theta \lesssim -0.01$, $127\,\text{GeV}\lesssim
m_{H}\lesssim 190\,\text{GeV}$, 
\begin{equation}
\begin{array}{llll}
\text{Br}(\tau \rightarrow \mu \gamma ) & \in (0.3-1.2)\times 10^{-8}, & 
\qquad \text{Br}(\tau \rightarrow e\gamma ) & \in (0.3-1.2)\times 10^{-8},
\\ 
\text{Br}(\tau \rightarrow 3\mu ) & \in (2.0-7.6)\times 10^{-10}, & \qquad 
\text{Br}(\tau \rightarrow 3e) & \in (1.1-4.1)\times 10^{-9}, \\ 
\text{Br}(h\rightarrow e\tau ) & \in (0.5-1.4)\times 10^{-2}, & \qquad \text{%
Br}(h\rightarrow e\mu ) & \in (1.5-4.1)\times 10^{-5},%
\end{array}%
\end{equation}%
are throughout below the current bounds in these particular parameter
regions, but not too small to be out of reach for future experiments. A
measurement of the $h\rightarrow e\tau $ channel using the newest LHC data
should be the fastest way to rule out this model in the near future, as our
prediction for this channel is unambiguously connected to $h\rightarrow \mu
\tau $ and expected to be the same order of magnitude, 
\begin{equation*}
\frac{\text{Br}(h\rightarrow \mu \tau )}{\text{Br}(h\rightarrow e\tau )}%
\approx \frac{|y_{\mu \tau }|^{2}}{|y_{e\tau }|^{2}}\approx \frac{(c_{\alpha
_{L}}+s_{\alpha _{L}})^{2}}{(c_{\alpha _{L}}-s_{\alpha _{L}})^{2}},
\end{equation*}%
which is approximately 1 for small values of $\theta $. Using Eqs. (\ref%
{e:jnp}) and (\ref{e:jp}), we also predict a nonvanishing leptonic CP phase $%
\delta _{CP}^{L}\approx 0.3$ at $\theta =0.07$, the maximal value still in
agreement with the lepton mixing angle $\theta _{12}$.

\section{Constraints from $h\rightarrow\protect\gamma 
\protect\gamma $}
\label{s:hgg}

As shown in the previous section, small $Z_3$ breaking perturbations allow to
successfully accommodate the experimental CMS data on the $h\to\mu\tau$
excess. Consequently and in order to make definite predictions, it is
reasonable to neglect these perturbations in the computation of the Higgs
diphoton decay rate. In the SM, the $h\rightarrow \gamma \gamma $
decay is dominated by $W$-loop diagrams which can interfere destructively
with the subdominant top quark loop. In our 3HDM, the $%
h\rightarrow \gamma \gamma $ decay receives additional contributions from
loops with charged scalars $\phi _{(a,b)}^{\pm }$, as shown in Fig. \ref%
{fig:hgg}, and therefore sets bounds on the masses of these scalars. 
\begin{figure}[tbp]
\includegraphics[width=1.1\textwidth]{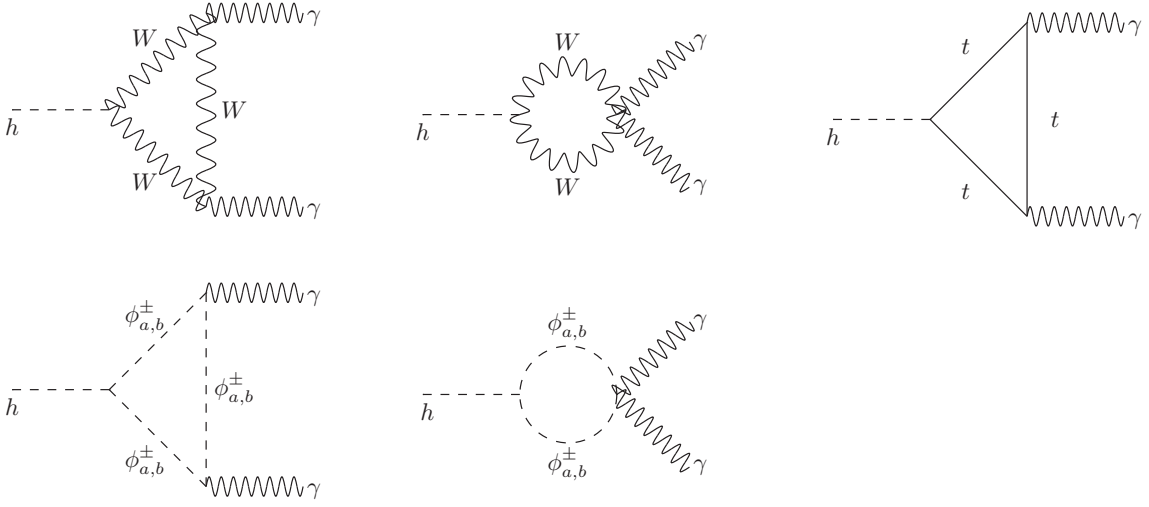}\vspace{-17cm}
\caption{One-loop Feynman diagrams in the unitary gauge contributing to the $%
h\rightarrow \protect\gamma \protect\gamma $ decay.}
\label{fig:hgg}
\end{figure}

The explicit form of the $h\rightarrow \gamma \gamma $ decay rate is \cite%
{Shifman:1979eb,Gavela:1981ri,Kalyniak:1985ct,Gunion:1989we,Spira:1997dg,Djouadi:2005gj}%
\begin{equation}
\Gamma \left( h\rightarrow \gamma \gamma \right) =\frac{\alpha
_{em}^{2}m_{h}^{3}}{256\pi ^{3}v^{2}}\left\vert
\sum_{f}a_{hff}N_{c}Q_{f}^{2}F_{1/2}\left( \beta _{f}\right) +F_{1}\left(
\beta _{W}\right) +\sum_{s=a,b}\frac{\lambda _{h\phi _{s}^{\pm }\phi
_{s}^{\mp }}v}{2m_{\phi _{s}^{\pm }}^{2}}F_{0}\left( \beta _{\phi _{s}^{\pm
}}\right) \right\vert ^{2} \label{e:hggcr}
\end{equation}
Here $\beta _{i}$ are the mass ratios $\beta _{i}=\frac{m_{h}^{2}}{4M_{i}^{2}%
}$, with $M_{i}=m_{f},M_{W}$, and $m_{\phi _{a,b}^{\pm }}$%
, $\alpha _{em}$ is the fine structure constant, $N_{C}$ is
the color factor ($N_{C}=1$ for leptons, $N_{C}=3$ for quarks), and $Q_{f}$
is the electric charge of the fermion in the loop. From the fermion-loop
contributions we consider only the dominant top quark term.
Furthermore, $\lambda _{h\phi _{s}^{\pm }\phi _{s}^{\mp }}$ is the trilinear
coupling between the SM-like Higgs and a pair of charged Higgses, which in
the case of an exact $Z_{3}$ symmetry is given by 
\begin{equation}
\lambda _{h\phi _{s}^{\pm }\phi _{s}^{\mp }}=\frac{2}{3}\left( 3\alpha
+2\beta -\gamma -\delta \right) v=\frac{2\left( m_{h}^{2}-m_{\phi _{s}^{\pm
}}^{2}\right) }{v},\qquad s=a,b.
\end{equation}
The dimensionless loop factors $F_{1/2}\left( \beta \right) $ and $%
F_{1}\left( \beta \right)$ can be found in Eqs. (\ref{e:f12}) - (\ref{e:fcase}) of Appendix \ref{a:hgg}.

In what follows we determine the range of values for the
charged Higgs boson masses $m_{\phi_{a,b}^{\pm }}$ which are consistent with
the $h\rightarrow \gamma \gamma $ results at the LHC. To this end, we
introduce the ratio $R_{\gamma \gamma }$, which normalizes the $\gamma\gamma$
signal predicted by our model relative to that of the SM: 
\begin{eqnarray}
R_{\gamma \gamma }&=&\frac{\sigma\left(pp\rightarrow h \right)\Gamma \left(
h\rightarrow \gamma \gamma \right) }{\sigma\left(pp\rightarrow h
\right)_{SM}\Gamma \left( h\rightarrow \gamma \gamma \right) _{SM}}\simeq
a^2_{htt}\frac{\Gamma \left( h\rightarrow \gamma \gamma \right) }{\Gamma
\left( h\rightarrow \gamma \gamma \right) _{SM}},  \label{R_gamma}
\end{eqnarray}
where $a_{htt}$ is the deviation of the Higgs-top quark coupling with
respect to the SM. Here we set $a_{htt}$ to be equal to one as in the
SM based on the fact that in our model single Higgs production is also dominated by
gluon fusion. The normalization given by Eq. (\ref{R_gamma}) for $%
h\rightarrow \gamma \gamma$ was also used in Refs.~\cite%
{Wang:2012gm,Carcamo-Hernandez:2013ypa,Hernandez:2015xka}.

\begin{figure}[tbp]
\includegraphics[width=0.8\textwidth]{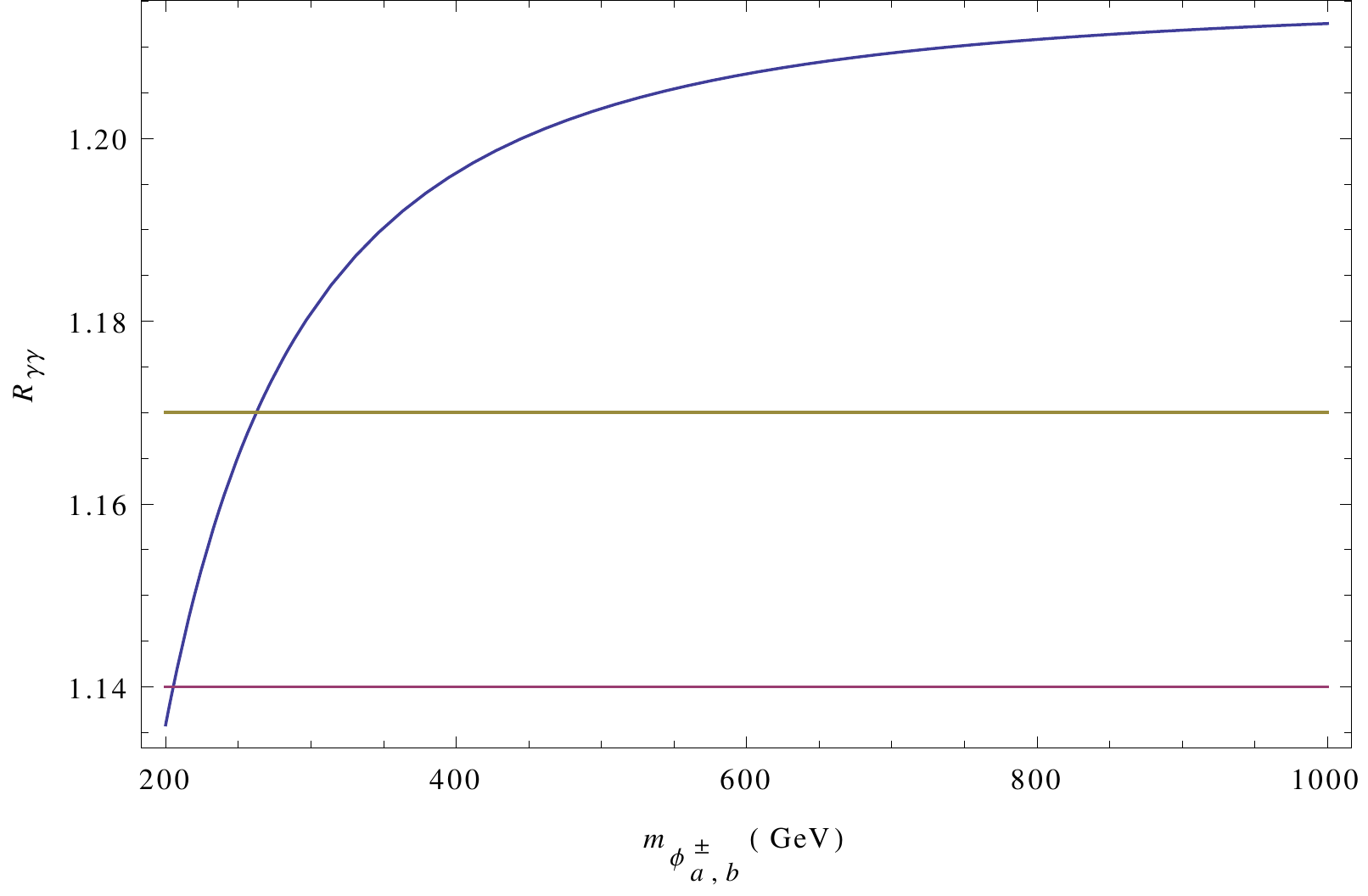}
\caption{The ratio $R_{\protect\gamma \protect\gamma }$ as a function of the
charged Higgs masses $m_{\protect\phi _{s}^{\pm }}$ ($s=a,b$) for $%
a_{htt}=1 $. The horizontal lines are the $R_{\protect\gamma \protect\gamma %
} $ experimental values given by CMS and ATLAS, which are equal to $%
1.14_{-0.23}^{+0.26}$ and $1.17\pm 0.27$, respectively \protect\cite%
{Khachatryan:2014ira,Aad:2014eha}.} 
\label{fig1}
\end{figure}

The ratio $R_{\gamma\gamma}$ has been measured by CMS and ATLAS with 
the best-fit signals \cite{Khachatryan:2014ira,Aad:2014eha}
\begin{align}
R_{\gamma\gamma}^{\text{CMS}}= 1.14_{-0.23}^{+0.26}\qquad \text{and} \qquad R_{\gamma\gamma}^{\text{ATLAS}}= 1.17\pm 0.27,
\end{align}
which are used to limit the charged scalar masses contributing to $h \rightarrow \gamma \gamma$. 
Figure \ref{fig1} shows the sensitivity of the ratio $R_{\gamma \gamma }$ under
variations of the charged Higgs masses $m_{\phi_{s}^{\pm }}$ ($s=a,b$),
between $200$ GeV and $1$ TeV. We consider charged Higgs boson masses larger 
than $200$ GeV to ensure they
are in the region above the $W^{\pm}h$ threshold, as done in Ref \cite%
{Enberg:2015swa}. The ratio $R_{\gamma \gamma }$ slowly increases when the
charged Higgs boson masses are increased. Requiring that the $h\to
\gamma\gamma$ signal stays within the range $1.14\lesssim R_{\gamma \gamma
}\lesssim 1.17$ (the central values of the recent CMS
and ATLAS results, respectively), yields the bound $200$ GeV$\lesssim
m_{\phi _{a,b}^{\pm }}\lesssim 205$ GeV for the charged Higgs boson masses.
However, considering the experimental errors of the measurements, 
the parameter space consistent with the Higgs diphoton signal becomes larger, i.e., masses in the range $200$ GeV$\lesssim
m_{\phi _{a,b}^{\pm }}\lesssim 1$ TeV can successfully account for the $h\to\gamma\gamma$ rate observed at the LHC. 
The scalar sector can be constrained further from the analysis of the $T$ and $S$ parameters, for details see Appendix \ref{a:TandS}.

\section{Conclusions}
\label{s:summ}
In summary we have presented a 3HDM with LFT, where strongly constrained FCNCs are suppressed by virtue of a residual $Z_3$ symmetry.
A small breaking of the $Z_3$ symmetry can give rise to LFV Higgs decays, occuring naturally in our model due to $SU(2)$ singlet scalars in the scalar potential. Furthermore, a small breaking of the $Z_3$ symmetry also gives rise to CP violation in neutrino oscillations.
We obtain a sizable branching ratio particularly in the flavor violating $h
\rightarrow \mu \tau^c$ channel, accounting for the $2.4\,\sigma$ deviation from the SM as indicated by recent CMS results.
This is a consequence of mixing between an SM-like Higgs with a nonstandard neutral scalar that
can decay into flavor violating final states. If the extra neutral scalars are light, they can partly cancel Higgs loop contributions to $l \rightarrow l^{\prime} \gamma$, allowing for large flavor violating Yukawa couplings. We thus also expect large branching fractions for $h \rightarrow e \mu$ and $h \rightarrow e \tau$ in
our model, where the latter will be a decisive measurement for its exclusion.
Furthermore, the $h \rightarrow \gamma \gamma$ rate measured by ATLAS and CMS favors charged Higgs boson masses of less than 205 GeV, but the large experimental uncertainties still allow for masses of up to 1 TeV in our model.

\section*{Acknowledgments}

We thank Gautam Bhattacharyya and Ivo de Medeiros Varzielas for useful
discussions. A.E.C.H thanks the TU Dortmund for hospitality and for
partially financing his visit. A.E.C.H was partially supported by Fondecyt
(Chile), Grant No. 11130115 and by DGIP internal Grant No. 111458. M.D.C.
thanks the DAAD for supporting his visit to the TU Dortmund. 
\vspace{-0.2cm} 

\newpage 
\begin{appendix}

 \section{$S_4$ Symmetry}
\label{a:rules}
$S_4$, the group of permutations of four objects, contains five irreducible representations ${\bf 1,1',2,3,3'}$ obeying the following tensor product rules \cite{Ishimori:2010au}:
\begin{align}
 &\mathbf{3} \otimes \mathbf{3} = \mathbf{1} \oplus \mathbf{2} \oplus \mathbf{3} \oplus \mathbf{3'}, \qquad \mathbf{3'} \otimes \mathbf{3'} = \mathbf{1} \oplus \mathbf{2} \oplus \mathbf{3} \oplus \mathbf{3'}, \qquad \mathbf{3} \otimes \mathbf{3'} = \mathbf{1'} \oplus \mathbf{2} \oplus \mathbf{3} \oplus \mathbf{3'}, \\
 &\mathbf{2} \otimes \mathbf{2} = \mathbf{1} \oplus \mathbf{1'} \oplus \mathbf{2}, \qquad \mathbf{2} \otimes \mathbf{3} = \mathbf{3} \oplus \mathbf{3'}, \qquad \mathbf{2} \otimes \mathbf{3'} = \mathbf{3'} \oplus \mathbf{3}, \\
 &\mathbf{3} \otimes \mathbf{1'} = \mathbf{3'}, \qquad \mathbf{3'} \otimes \mathbf{1'} = \mathbf{3}, \qquad \mathbf{2} \otimes \mathbf{1'} = \mathbf{2}.
\end{align}
Explicitly, the basis used in this paper corresponds to Ref. \cite{Ishimori:2010au} and results in
\begin{eqnarray}
 &({\bf A})_{\bf3}\times({\bf B})_{\bf3}=({\bf A}\cdot{\bf B})_{\bf1}
+ \left(\begin{array}{c} {\bf A}\cdot\Sigma\cdot{\bf B}  \\ {\bf A}\cdot\Sigma^*\cdot{\bf B}  \end{array} \right)_{\bf2}
+\left(\begin{array}{c} \{A_yB_z\} \\ \{A_zB_x\} \\ \{A_xB_y\} \end{array} \right)_{\bf3}
+\left(\begin{array}{c} \left[A_yB_z\right] \\ \left[A_zB_x\right] \\ \left[A_xB_y\right] \end{array} \right)_{\bf3'} , \end{eqnarray}
\begin{eqnarray}
&({\bf A})_{\bf3'}\times({\bf B})_{\bf3'}=({\bf A}\cdot{\bf B})_{\bf1}
+\left(\begin{array}{c} {\bf A}\cdot\Sigma\cdot{\bf B} \\  {\bf A}\cdot\Sigma^*\cdot{\bf B} \end{array} \right)_{\bf2}
+\left(\begin{array}{c} \{A_yB_z\} \\ \{A_zB_x\} \\ \{A_xB_y\} \end{array} \right)_{\bf3}
+\left(\begin{array}{c} \left[A_yB_z\right] \\ \left[A_zB_x\right] \\ \left[A_xB_y\right] \end{array} \right)_{\bf3'}, \end{eqnarray}
\begin{eqnarray}
&({\bf A})_{\bf3}\times({\bf B})_{\bf3'}=({\bf A}\cdot{\bf B})_{\bf1'}
+\left(\begin{array}{c} {\bf A}\cdot\Sigma\cdot{\bf B} \\ -{\bf A}\cdot\Sigma^*\cdot{\bf B}  \end{array} \right)_{\bf2}
+\left(\begin{array}{c} \{A_yB_z\} \\ \{A_zB_x\} \\ \{A_xB_y\} \end{array} \right)_{\bf3'}
+\left(\begin{array}{c} \left[A_yB_z\right] \\ \left[A_zB_x\right] \\ \left[A_xB_y\right] \end{array} \right)_{\bf3}, \end{eqnarray}
\begin{eqnarray}
&({\bf A})_{\bf2}\times({\bf B})_{\bf2}=
\{A_xB_y\}_{\bf1}+\left[A_xB_y\right]_{\bf1'}
+\left(\begin{array}{c} A_yB_y \\ A_xB_x \end{array} \right)_{\bf2} , \end{eqnarray}
\begin{eqnarray}
&\left(\begin{array}{c} A_x \\ A_y \end{array} \right)_{\bf 2}\times \left(\begin{array}{c} B_x \\ B_y \\ B_z \end{array} \right)_{\bf 3}
=\left(\begin{array}{c} (A_x+A_y)B_x \\ (\omega^2A_x+\omega A_y)B_y \\ (\omega
  A_x+\omega^2 A_y)B_z \end{array} \right)_{\bf 3}+ 
\left(\begin{array}{c} (A_x-A_y)B_x \\ (\omega^2A_x-\omega A_y)B_y \\ (\omega A_x-\omega^2
  A_y)B_z \end{array} \right)_{{\bf 3}'}, \end{eqnarray}
\begin{eqnarray}
&\left(\begin{array}{c} A_x \\ A_y \end{array} \right)_{\bf 2}\times \left(\begin{array}{c} B_x \\ B_y \\ B_z \end{array} \right)_{{\bf 3}'}
=\left(\begin{array}{c}(A_x+A_y)B_x \\ (\omega^2A_x+\omega A_y)B_y \\ (\omega
  A_x+\omega^2 A_y)B_z \end{array} \right)_{{\bf 3}'} + 
\left(\begin{array}{c}(A_x-A_y)B_x \\ (\omega^2A_x-\omega A_y)B_y \\ (\omega A_x-\omega^2
  A_y)B_z \end{array} \right)_{\bf 3},
\end{eqnarray}
with
\begin{align}
{\bf A}\cdot{\bf B}&=A_xB_x+A_yB_y+A_zB_z,   \\
\{A_x B_y\}&=A_x B_y + A_y B_x,  \\
\left[A_x B_y\right]&= A_x B_y - A_y B_x, \\
{\bf A}\cdot\Sigma\cdot{\bf B} &=A_xB_x+\omega A_yB_y+\omega^2A_zB_z,
 \\
{\bf A}\cdot\Sigma^*\cdot{\bf B} &=A_xB_x+\omega^2 A_yB_y+\omega
A_zB_z , 
\end{align}
where $\omega=e^{2\pi i/3}$ is a complex square root of unity.

\section{T and S parameter constraints}
\label{a:TandS}

The inclusion of the extra scalar particles modifies the oblique corrections
of the SM, the values which have been extracted from high precision
experiments. Consequently, the validity of the different flavor models that
we are considering depends on the condition that the extra particles do not
contradict those experimental results. These oblique corrections are
parametrized in terms of the two well-known quantities $T$ and $S$. The $T$
parameter is defined as \cite%
{Peskin:1991sw,Altarelli:1990zd,Barbieri:2004qk,Barbieri:2007gi}: 
\begin{equation}
T=\frac{\Pi _{33}\left( 0\right) -\Pi _{11}\left( 0\right) } {M_{W}^{2} \
\alpha _{em}\left( m_{Z}\right) }.
\end{equation}
where $\Pi _{11}\left( 0\right) $ and $\Pi _{33}\left( 0\right) $ are the
vacuum polarization amplitudes at $q^2 =0$ for loop diagrams having gauge
bosons $W_{\mu }^{1}$, $W_{\mu }^{1}$ and $W_{\mu }^{3}$, $W_{\mu }^{3}$ in
the external lines, respectively.

In turn, the $S$ parameter is defined by \cite{Peskin:1991sw,Altarelli:1990zd,Barbieri:2004qk,Barbieri:2007gi}: 
\begin{equation}
S=\frac{4\sin ^{2}\theta _{W}}{\alpha_{em}\left( m_{Z}\right) } \frac{g}{%
g^{\prime }}\frac{d\Pi _{30}\left( q^{2}\right) }{dq^{2}}\biggl|_{q^{2}=0},
\end{equation}
where $\Pi _{30}\left( q^{2}\right) $ is the vacuum polarization amplitude
for a loop diagram having $W_{\mu }^{3}$ and $B_{\mu }$ in the external
lines.

\begin{figure}[tbh]
\includegraphics[width=0.7\textwidth]{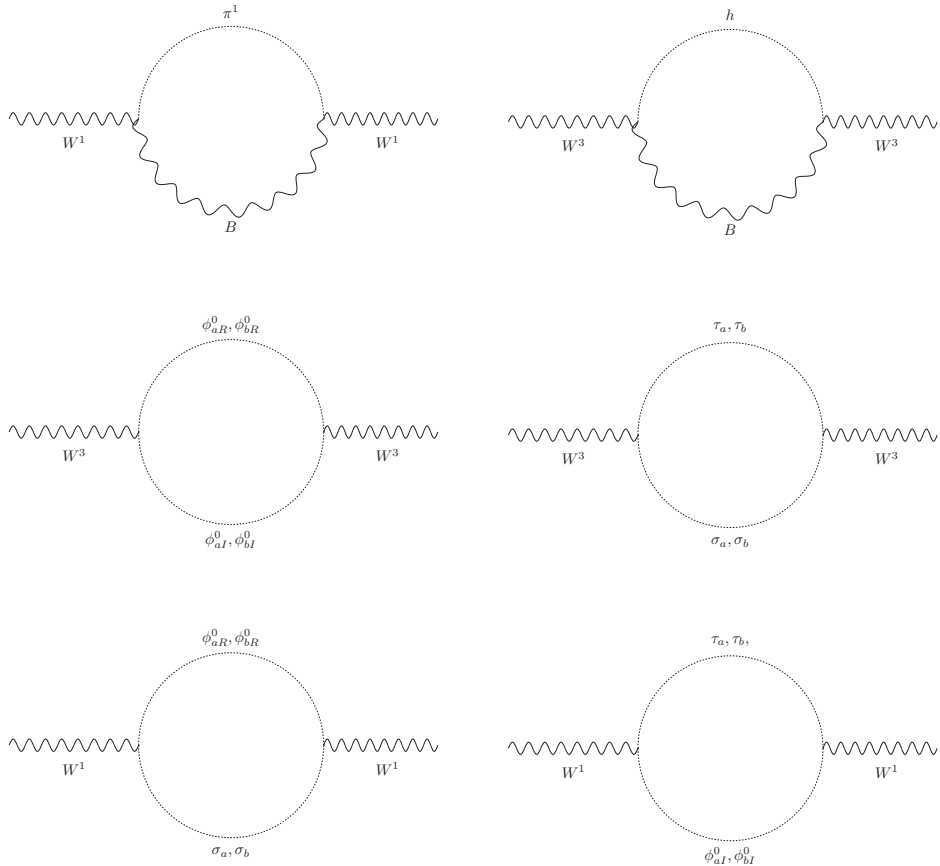}\vspace{-3cm}
\caption{One-loop Feynman diagrams contributing to the $T$ parameter.}
\label{diagT1}
\end{figure}

\begin{figure}[tbh]
\vspace{-3cm} \includegraphics[width=0.7\textwidth]{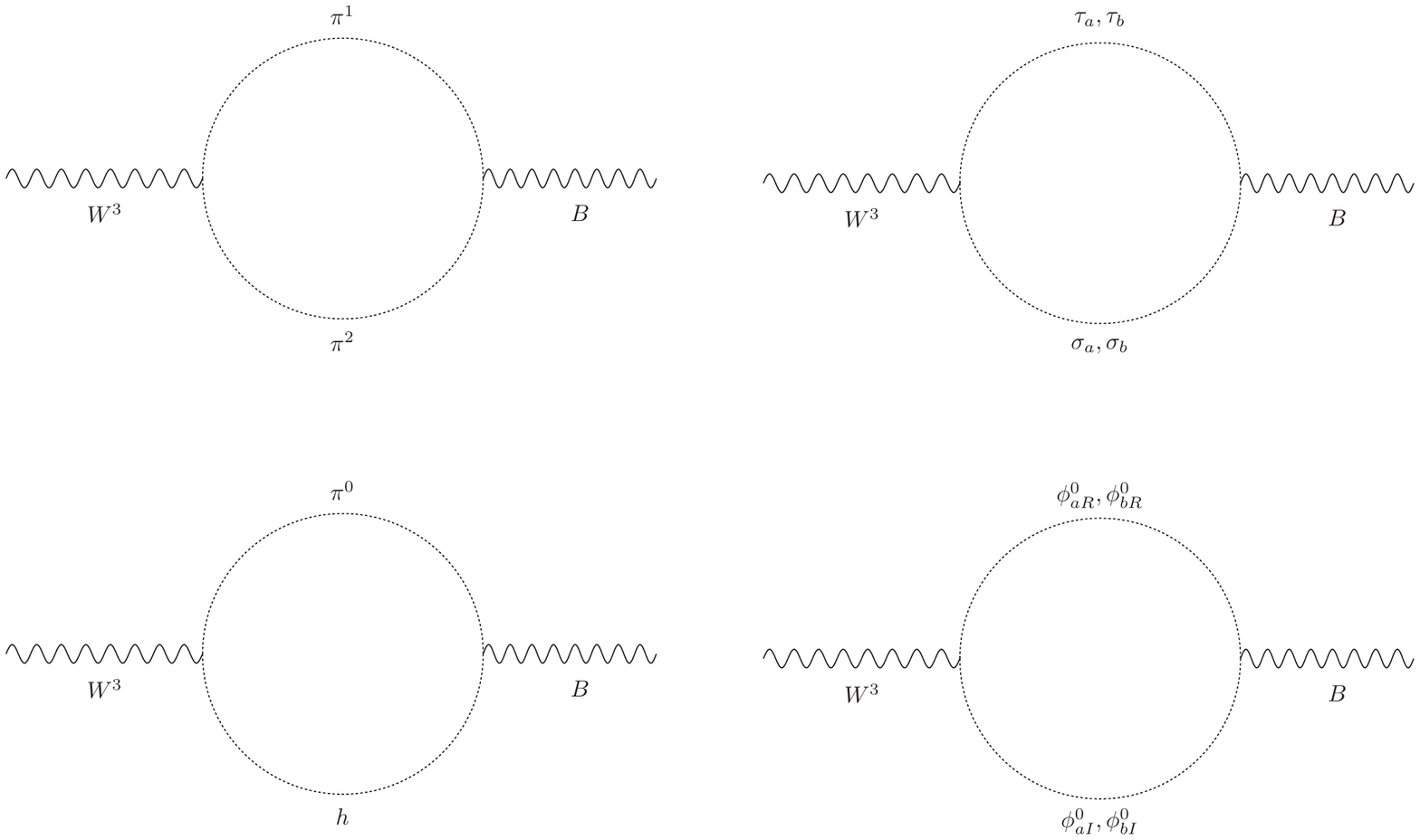}\vspace{%
-5cm}
\caption{One-loop Feynman diagrams contributing to the $S$ parameter.}
\label{diagS1}
\end{figure}

The Feynman diagrams contributing to the $T$ and $S$ parameters are shown in
Figs. \ref{diagT1} and \ref{diagS1}. 


We can split the $T$ and $S$ parameters as $T=T_{SM}+\Delta T$ and $%
S=S_{SM}+\Delta S$, where $T_{SM}$ and $S_{SM}$\ are the contributions from
the SM, while $\Delta T$ and $\Delta S$ contain all the contributions
involving the extra particles 
\begin{equation}
T_{SM}=-\frac{3}{16\pi \cos ^{2}\theta _{W}}\ln \left( \frac{m_{h}^{2}}{%
m_{W}^{2}}\right) ,\hspace{0.2cm}\hspace{0.2cm}\hspace{0.2cm}\hspace{0.2cm}%
\hspace{0.2cm}\hspace{0.2cm}\hspace{0.2cm}S_{SM}=\frac{1}{12\pi }\ln \left( 
\frac{m_{h}^{2}}{m_{W}^{2}}\right)  \label{St}
\end{equation}
\begin{equation}
\Delta T\simeq \frac{1}{16\pi ^{2}v^{2}\alpha _{em}\left( m_{Z}\right) }%
\sum_{f=a,b}\left[ m_{m_{\phi _{f}^{\pm }}^{2}}^{2}-H\left( m_{\phi
_{fI}^{0}}^{2},m_{\phi _{f}^{\pm }}^{2}\right) +H\left( m_{\phi
_{fR}^{0}}^{2},m_{\phi _{fI}^{0}}^{2}\right) -H\left( m_{\phi
_{fR}^{0}}^{2},m_{\phi _{f}^{\pm }}^{2}\right) \right]
\end{equation}
\begin{equation}
\Delta S\simeq \frac{1}{12\pi }\sum_{f=a,b}K\left( m_{\phi
_{fR}^{0}}^{2},m_{\phi _{fI}^{0}}^{2},m_{\phi _{f}^{\pm }}^{2}\right) ,
\end{equation}%
where we introduced the following functions \cite{Hernandez:2015rfa}:
\begin{equation}
H\left( m_{1}^{2},m_{2}^{2}\right) =\frac{m_{1}^{2}m_{2}^{2}}{%
m_{1}^{2}-m_{2}^{2}}\ln \left( \frac{m_{1}^{2}}{m_{2}^{2}}\right) ,\hspace{%
1.5cm}\hspace{1.5cm}\lim_{m_{2}\rightarrow m_{1}}H\left(
m_{1}^{2},m_{2}^{2}\right) =m_{1}^{2}.
\end{equation}%
\begin{eqnarray}
K\left( m_{1}^{2},m_{2}^{2},m_{3}^{2}\right) &=&\frac{1}{\left(
m_{2}^{2}-m_{1}^{2}\right) {}^{3}}\left\{ m_{1}^{4}\left(
3m_{2}^{2}-m_{1}^{2}\right) \ln \left( \frac{m_{1}^{2}}{m_{3}^{2}}\right)
-m_{2}^{4}\left( 3m_{1}^{2}-m_{2}^{2}\right) \ln \left( \frac{m_{2}^{2}}{%
m_{3}^{2}}\right) \right.  \notag \\
&&-\left. \frac{1}{6}\left[ 27m_{1}^{2}m_{2}^{2}\left(
m_{1}^{2}-m_{2}^{2}\right) +5\left( m_{2}^{6}-m_{1}^{6}\right) \right]
\right\} .
\end{eqnarray}

\begin{figure}[tbh]
\subfigure[~$m_{\phi_{(a,b)I}^{0}}=m_{\phi_{(a,b)R}^{0}}=130$
GeV]{\includegraphics[width=0.32\textwidth]{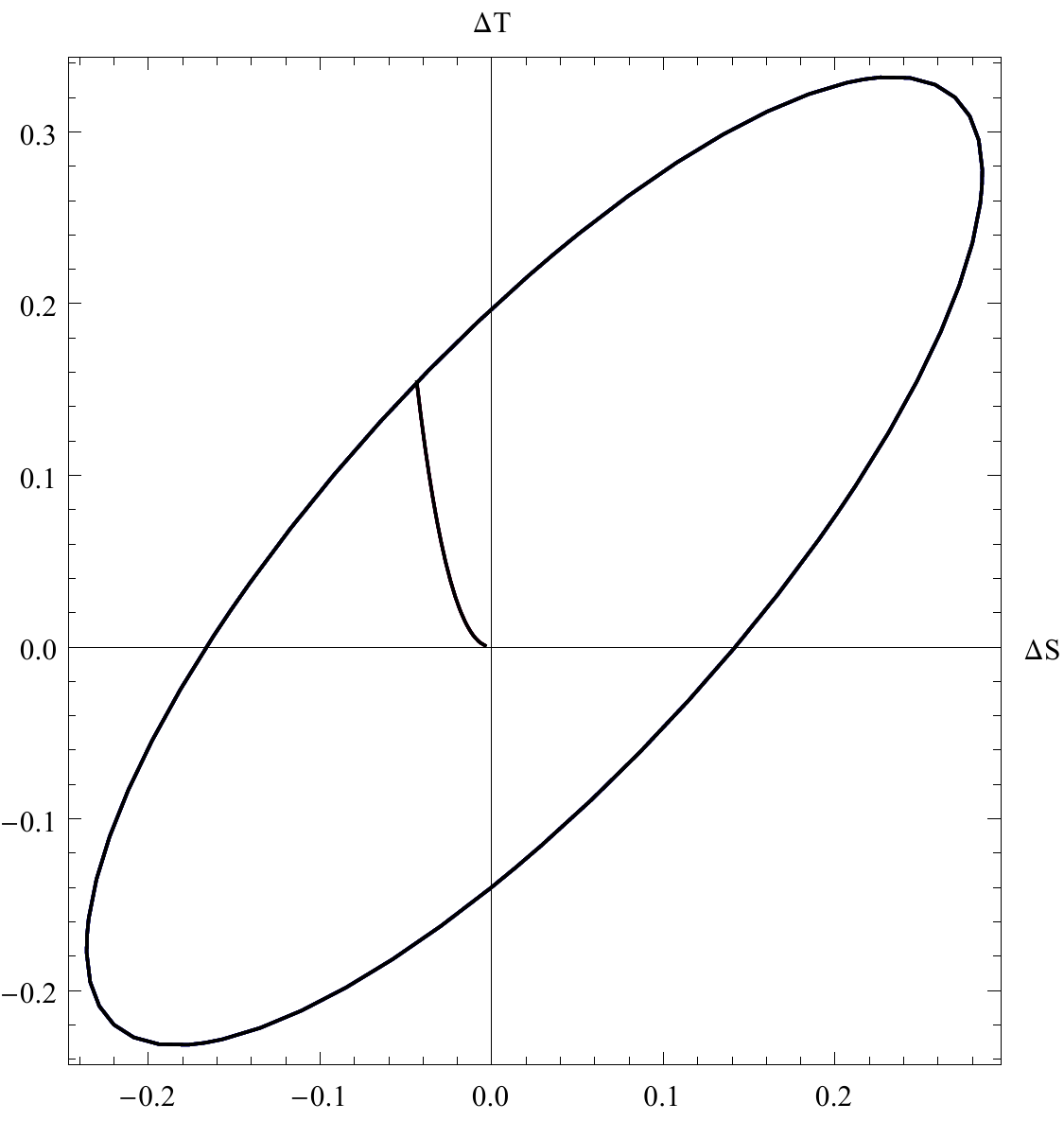}} 
\subfigure[~$m_{\phi_{(a,b)I}^{0}}=m_{\phi_{(a,b)R}^{0}}=600$
GeV]{\includegraphics[width=0.32\textwidth]{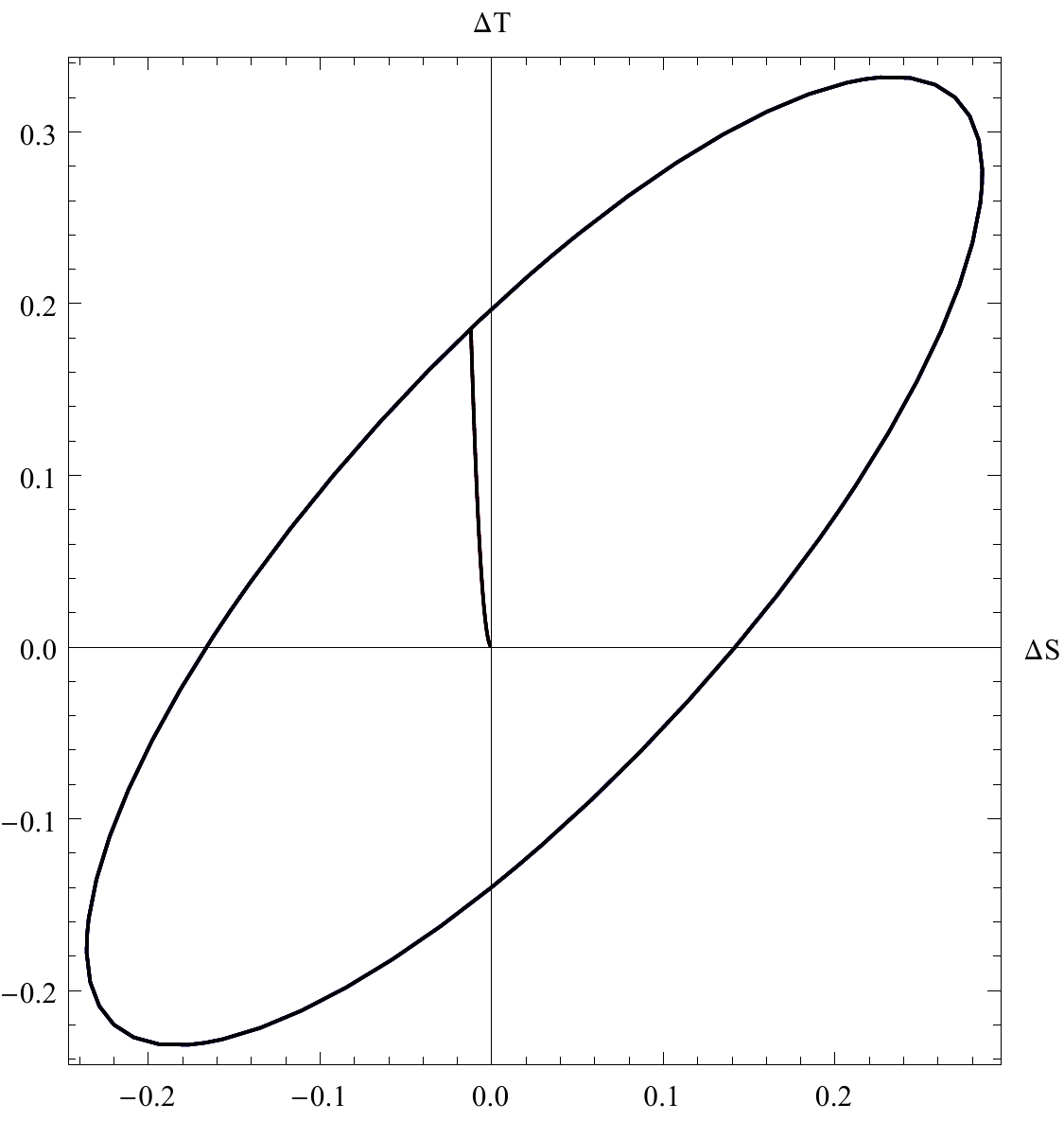}} 
\subfigure[~$m_{\phi_{(a,b)I}^{0}}=m_{\phi_{(a,b)R}^{0}}=1$
TeV]{\includegraphics[width=0.32\textwidth]{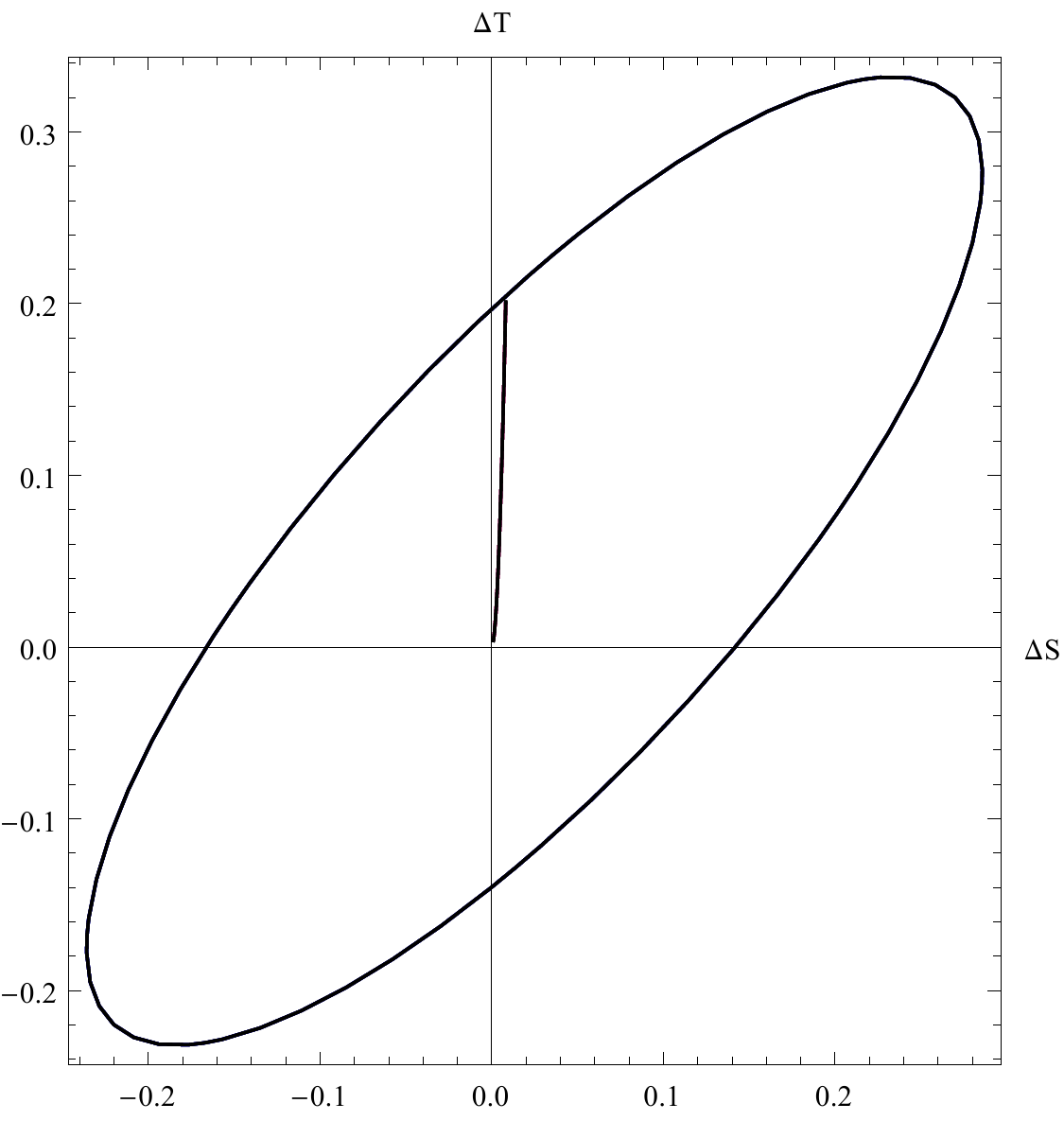}} 
{\footnotesize {\ \ \ (\ref{TandS1}.a)}}\hspace{5.5cm}{\footnotesize {(\ref%
{TandS1}.b)}}\hspace{5.5cm}{\footnotesize {(\ref{TandS1}.c)}} \newline
\caption{The $\Delta S-\Delta T$ plane, where the ellipses denote the
experimentally allowed region at $95\%$ C.L. taken from \protect\cite%
{Baak:2013ppa,Baak:2012kk,Baak:2011ze}. The origin $\Delta S=\Delta T=0$
corresponds to the SM value, with $m_{h}=125.5$~GeV and $%
m_{t}=176$~GeV. Figures (a), (b) and (c) correspond to two different sets of values
for the masses of the neutral non-SM Higgs bosons, as indicated. The mass $%
m_{\protect\phi_{(a,b)}^{\pm}}$ of the charged Higgs bosons varies between $%
130$ GeV $\leq m_{\protect\phi_{(a,b)}^{\pm}}\leq$ $196$ GeV (Fig. \ref{TandS1}%
(a), $600$ GeV $\leq m_{\protect\phi_{(a,b)}^{\pm}}\leq$ $672$ GeV ( 
Fig. \ref{TandS1}(b), $925$ GeV $\leq m_{\protect\phi_{(a,b)}^{\pm}}\leq$ $990$
GeV (Fig. \ref{TandS1}(c). The lines originating in the center of the plot and running up towards the ellipses correspond to the obtained values of the $\Delta T$ and $\Delta S$ parameters in our model, as the charged Higgs boson masses are varied in the aforementioned ranges. }
\label{TandS1}
\end{figure}

The experimental results on $T$ and $S$ restrict $\Delta T$ and $\Delta S$
to lie inside a region in the $\Delta S-\Delta T$ plane. At the $95\%$ C.L.
(confidence level), these regions are the elliptic contours shown in %
Fig. \ref{TandS1}. The origin $\Delta S=\Delta T=0$ corresponds to the SM value, 
with $m_{h}=125.5$~GeV and $m_{t}=176$~GeV. 
One can consider a scenario in which the heavy neutral CP-even
and neutral CP-odd scalars are degenerate. For a mass of $m_{\phi
_{(a,b)I}^{0}}=m_{\phi _{(a,b)R}^{0}}=130$ GeV, consistency with the $T$ and 
$S$ parameters confines the masses of the charged scalars to $130$ GeV $\leq
m_{\phi _{(a,b)}^{\pm }}\leq $ $196$ GeV, whereas for $m_{\phi
_{(a,b)I}^{0}}=m_{\phi _{(a,b)R}^{0}}=600$ GeV, the charged scalar masses
are in the range $600$ GeV $\leq m_{\phi _{(a,b)}^{\pm }}\leq $ $672$ GeV\
while for $m_{\phi _{(a,b)I}^{0}}=m_{\phi _{(a,b)R}^{0}}=1$ TeV, the masses
of the charged scalars range between 925 and 990 GeV. 
In Figs. \ref{TandS1}(a)-7(c) we show the allowed regions for
the $\Delta T$ and $\Delta S$ parameters, for the three sets of values of $%
m_{\phi_{(a,b)I}^{0}}$ and $m_{\phi_{(a,b)R}^{0}}$ previously indicated. 
The line
going upwards inside the ellipses corresponds to the several $(\Delta
S,\Delta T)$ points of the model when the charged scalar masses are varied
inside the interval previously specified. 

\section{Quark masses and mixings}
\label{a:quarks}
The relevant $S_{4}\otimes Z_{2}^{\prime \prime }\otimes Z_{6}\otimes Z_{12}$
-invariant Yukawa terms for the up-type quark sector are
\begin{eqnarray}
\mathcal{L}^{\left( U\right) } &=&y_{1}^{\left( t\right) }\left[ Q\phi %
\right] _{3^{\prime }}t_{R}\frac{\rho }{\Lambda }+\frac{y_{2}^{\left(
t\right) }}{\Lambda }Q\left[ \phi \rho \right] _{3^{\prime
}}t_{R}+y_{3}^{\left( t\right) }\left[ Q\phi \right] _{3^{\prime }}t_{R}%
\frac{\rho }{\Lambda }+\frac{y_{4}^{\left( t\right) }}{\Lambda }Q\left[ \phi
\rho \right] _{3^{\prime }}t_{R}  \notag \\
&&+x_{1}^{\left( t\right) }\left[ Q\phi \right] _{3}t_{R}\frac{\varphi }{%
\Lambda }+\frac{x_{2}^{\left( t\right) }}{\Lambda }Q\left[ \phi \varphi %
\right] _{3^{\prime }}t_{R}+x_{3}^{\left( t\right) }\left[ Q\phi \right]
_{3}t_{R}\frac{\varphi }{\Lambda }+\frac{x_{4}^{\left( t\right) }}{\Lambda }Q%
\left[ \phi \varphi \right] _{3^{\prime }}t_{R}  \notag \\
&&+y_{1}^{\left( c\right) }\left[ Q\phi \right] _{3^{\prime }}c_{R}\frac{%
\rho }{\Lambda }+\frac{y_{2}^{\left( c\right) }}{\Lambda }Q\left[ \phi \rho %
\right] _{3^{\prime }}c_{R}+y_{3}^{\left( c\right) }\left[ Q\phi \right]
_{3^{\prime }}c_{R}\frac{\rho }{\Lambda }+\frac{y_{4}^{\left( c\right) }}{%
\Lambda }Q\left[ \phi \rho \right] _{3^{\prime }}c_{R}  \notag \\
&&+x_{1}^{\left( c\right) }\left[ Q\phi \right] _{3}c_{R}\frac{\varphi }{%
\Lambda }+\frac{x_{2}^{\left( c\right) }}{\Lambda }Q\left[ \phi \varphi %
\right] _{3^{\prime }}c_{R}+x_{3}^{\left( c\right) }\left[ Q\phi \right]
_{3}c_{R}\frac{\varphi }{\Lambda }+\frac{x_{4}^{\left( c\right) }}{\Lambda }Q%
\left[ \phi \varphi \right] _{3^{\prime }}c_{R}+x_{0}^{\left( u\right) }%
\left[ Q\phi \right] _{1}u_{R}\frac{\Omega _{2}^{3}}{\Lambda ^{3}}  \notag \\
&&+y_{1}^{\left( u\right) }\left[ Q\phi \right] _{3^{\prime }}u_{R}\frac{%
\rho \Omega _{1}^{2}}{\Lambda ^{3}}+y_{2}^{\left( u\right) }Q\left[ \phi \rho %
\right] _{3^{\prime }}u_{R}\frac{\Omega _{1}^{2}}{\Lambda ^{3}}+y_{3}^{\left(
u\right) }\left[ Q\phi \right] _{3^{\prime }}u_{R}\frac{\rho \Omega _{1}^{2}}{%
\Lambda ^{4}}+\frac{y_{4}^{\left( u\right) }}{\Lambda }Q\left[ \phi \rho %
\right] _{3^{\prime }}u_{R}\frac{\Omega _{1}^{2}}{\Lambda ^{3}}  \notag \\
&&+x_{1}^{\left( u\right) }\left[ Q\phi \right] _{3^{\prime }}u_{R}\frac{%
\varphi \Omega _{1}^{2}}{\Lambda ^{3}}+x_{2}^{\left( u\right) }Q\left[ \phi
\varphi \right] _{3^{\prime }}u_{R}\frac{\Omega _{1}^{2}}{\Lambda ^{3}}%
+x_{3}^{\left( c\right) }\left[ Q\phi \right] _{3}u_{R}\frac{\varphi }{%
\Lambda }+\frac{x_{4}^{\left( c\right) }}{\Lambda }Q\left[ \phi \varphi %
\right] _{3^{\prime }}u_{R}\frac{\Omega _{1}^{2}}{\Lambda ^{3}}  \label{lyu}
\end{eqnarray}
and for the down-type quarks
\begin{eqnarray}
\mathcal{L}_{1}^{\left( D\right) } &=&y_{1}^{\left( b\right) }\left[ Q\phi %
\right] _{3^{\prime }}b_{R}\frac{\rho \Omega _{3}^{3}}{\Lambda ^{4}}+\frac{%
y_{2}^{\left( b\right) }}{\Lambda }Q\left[ \phi \rho \right] _{3^{\prime
}}b_{R}\frac{\Omega _{3}^{3}}{\Lambda ^{3}}+y_{3}^{\left( b\right) }\left[
Q\phi \right] _{3^{\prime }}b_{R}\frac{\rho \Omega _{3}^{3}}{\Lambda ^{4}}+%
\frac{y_{4}^{\left( b\right) }}{\Lambda }Q\left[ \phi \rho \right]
_{3^{\prime }}b_{R}\frac{\Omega _{3}^{3}}{\Lambda ^{3}}  \notag \\
&&+x_{1}^{\left( b\right) }\left[ Q\phi \right] _{3}b_{R}\frac{\varphi \Omega
_{3}^{3}}{\Lambda ^{4}}+\frac{x_{2}^{\left( b\right) }}{\Lambda }Q\left[
\phi \varphi \right] _{3^{\prime }}b_{R}\frac{\Omega _{3}^{3}}{\Lambda ^{3}}%
+x_{3}^{\left( b\right) }\left[ Q\phi \right] _{3}b_{R}\frac{\varphi \Omega
_{3}^{3}}{\Lambda ^{4}}+\frac{x_{4}^{\left( b\right) }}{\Lambda }Q\left[
\phi \varphi \right] _{3^{\prime }}b_{R}\frac{\Omega _{3}^{3}}{\Lambda ^{3}} 
\notag \\
&&+y_{1}^{\left( s\right) }\left[ Q\phi \right] _{3^{\prime }}s_{R}\frac{%
\rho \Omega _{3}^{3}}{\Lambda ^{4}}+\frac{y_{2}^{\left( b\right) }}{\Lambda }Q%
\left[ \phi \rho \right] _{3^{\prime }}s_{R}\frac{\Omega _{3}^{3}}{\Lambda ^{3}%
}+y_{3}^{\left( b\right) }\left[ Q\phi \right] _{3^{\prime }}s_{R}\frac{\rho
\Omega _{3}^{3}}{\Lambda ^{4}}+\frac{y_{4}^{\left( b\right) }}{\Lambda }Q\left[
\phi \rho \right] _{3^{\prime }}s_{R}\frac{\Omega _{3}^{3}}{\Lambda ^{3}} 
\notag \\
&&+x_{1}^{\left( s\right) }\left[ Q\phi \right] _{3}s_{R}\frac{\varphi \Omega
_{3}^{3}}{\Lambda ^{4}}+\frac{x_{2}^{\left( s\right) }}{\Lambda }Q\left[
\phi \varphi \right] _{3^{\prime }}s_{R}\frac{\Omega _{3}^{3}}{\Lambda ^{3}}%
+x_{3}^{\left( s\right) }\left[ Q\phi \right] _{3}s_{R}\frac{\varphi \Omega
_{3}^{3}}{\Lambda ^{4}}+\frac{x_{4}^{\left( s\right) }}{\Lambda }Q\left[
\phi \varphi \right] _{3^{\prime }}s_{R}\frac{\Omega _{3}^{3}}{\Lambda ^{3}} 
\notag \\
&&+x_{0}^{\left( d\right) }\left[ Q\phi \right] _{1}d_{R}\frac{\Omega
_{2}^{2}\Omega _{3}^{3}}{\Lambda ^{5}}+y_{1}^{\left( d\right) }\left[ Q\phi %
\right] _{3^{\prime }}d_{R}\frac{\rho \Omega _{1}\Omega _{3}^{3}}{\Lambda ^{5}}%
+y_{2}^{\left( d\right) }Q\left[ \phi \rho \right] _{3^{\prime }}d_{R}\frac{%
\Omega _{1}\Omega _{3}^{3}}{\Lambda ^{5}}+y_{3}^{\left( d\right) }\left[ Q\phi %
\right] _{3^{\prime }}d_{R}\frac{\rho \Omega _{1}\Omega _{3}^{3}}{\Lambda ^{5}} 
\notag \\
&&+\frac{y_{4}^{\left( d\right) }}{\Lambda }Q\left[ \phi \rho \right]
_{3^{\prime }}d_{R}\frac{\Omega _{1}\Omega _{3}^{3}}{\Lambda ^{5}}+x_{1}^{\left(
d\right) }\left[ Q\phi \right] _{3^{\prime }}d_{R}\frac{\varphi \Omega
_{1}\Omega _{3}^{3}}{\Lambda ^{5}}+x_{2}^{\left( d\right) }Q\left[ \phi
\varphi \right] _{3^{\prime }}d_{R}\frac{\Omega _{1}\Omega _{3}^{3}}{\Lambda ^{5}%
}  \label{ld} \\
&&+x_{3}^{\left( d\right) }\left[ Q\phi \right] _{3}d_{R}\frac{\varphi \Omega
_{3}^{3}}{\Lambda ^{5}}+x_{4}^{\left( d\right) }Q\left[ \phi \varphi \right]
_{3^{\prime }}d_{R}\frac{\Omega _{1}\Omega _{3}^{3}}{\Lambda ^{5}}  \notag.
\end{eqnarray}

\section{PMNS matrix after $Z_3$ breaking}
\label{a:mix}
The PMNS matrix receives corrections caused by the perturbation of the VEV alignment. These are approximately given by
\begin{eqnarray}
U &\simeq &\left( 
\begin{array}{ccc}
\frac{\cos \psi }{\sqrt{3}}-\frac{e^{i\phi -\frac{2i\pi }{3}}\sin \psi }{%
\sqrt{3}} & \frac{e^{\frac{2i\pi }{3}}}{\sqrt{3}} & \frac{e^{-\frac{2i\pi }{3%
}}\cos \psi }{\sqrt{3}}+\frac{e^{-i\phi }\sin \psi }{\sqrt{3}} \\ 
\frac{\cos \psi }{\sqrt{3}}-\frac{e^{i\phi +\frac{2i\pi }{3}}\sin \psi }{%
\sqrt{3}} & \frac{e^{-\frac{2i\pi }{3}}}{\sqrt{3}} & \frac{e^{\frac{2i\pi }{3%
}}\cos \psi }{\sqrt{3}}+\frac{e^{-i\phi }\sin \psi }{\sqrt{3}} \\ 
\frac{\cos \psi }{\sqrt{3}}-\frac{e^{i\phi }\sin \psi }{\sqrt{3}} & \frac{1}{%
\sqrt{3}} & \frac{\cos \psi }{\sqrt{3}}+\frac{e^{-i\phi }\sin \psi }{\sqrt{3}%
}%
\end{array}%
\right)  \notag \\
&+&\theta \left( 
\begin{array}{ccc}
-\sqrt{\frac{2}{3}}\cos \psi +\frac{e^{i\phi +\frac{2i\pi }{3}}\sin \psi }{%
\sqrt{6}}+\frac{e^{i\phi }\sin \psi }{\sqrt{6}} & -\frac{1}{\sqrt{6}}-\frac{%
e^{-\frac{2i\pi }{3}}}{\sqrt{6}} & -\frac{e^{\frac{2i\pi }{3}}\cos \psi }{%
\sqrt{6}}-\frac{\cos \psi }{\sqrt{6}}-\sqrt{\frac{2}{3}}e^{-i\phi }\sin \psi
\\ 
\frac{e^{i\phi }\sin \psi }{\sqrt{6}}-\frac{e^{i\phi -\frac{2i\pi }{3}}\sin
\psi }{\sqrt{6}} & -\frac{1}{\sqrt{6}}+\frac{e^{\frac{2i\pi }{3}}}{\sqrt{6}}
& \frac{e^{-\frac{2i\pi }{3}}\cos \psi }{\sqrt{6}}-\frac{\cos \psi }{\sqrt{6}%
} \\ 
\sqrt{\frac{2}{3}}\cos \psi -\frac{e^{i\phi -\frac{2i\pi }{3}}\sin \psi }{%
\sqrt{6}}-\frac{e^{i\phi +\frac{2i\pi }{3}}\sin \psi }{\sqrt{6}} & \frac{e^{-%
\frac{2i\pi }{3}}}{\sqrt{6}}+\frac{e^{\frac{2i\pi }{3}}}{\sqrt{6}} & \frac{%
e^{\frac{2i\pi }{3}}\cos \psi }{\sqrt{6}}+\frac{e^{-\frac{2i\pi }{3}}\cos
\psi }{\sqrt{6}}+\sqrt{\frac{2}{3}}e^{-i\phi }\sin \psi%
\end{array}%
\right)  \notag \\
&+&\theta ^{2}\left( 
\begin{array}{ccc}
\frac{1}{4}\sqrt{3}\cos \psi -\frac{1}{4}\sqrt{3}e^{i\phi +\frac{2i\pi }{3}%
}\sin \psi & \frac{1}{4}\sqrt{3}e^{-\frac{2i\pi }{3}} & \frac{1}{4}\sqrt{3}%
e^{\frac{2i\pi }{3}}\cos \psi +\frac{1}{4}\sqrt{3}e^{-i\phi }\sin \psi \\ 
\frac{1}{4}\sqrt{3}e^{i\phi -\frac{2i\pi }{3}}\sin \psi -\frac{1}{4}\sqrt{3}%
\cos \psi & -\frac{1}{4}\sqrt{3}e^{\frac{2i\pi }{3}} & -\frac{1}{4}\sqrt{3}%
e^{-\frac{2i\pi }{3}}\cos \psi -\frac{1}{4}\sqrt{3}e^{-i\phi }\sin \psi \\ 
0 & 0 & 0%
\end{array}%
\right),  \label{PMNSZ3B}
\end{eqnarray}
whereas the deviations in the mixing angles caused by the perturbation 
$\theta$ are accounted for by  
\begin{eqnarray}  \label{e:sin12p}
\sin ^{2}\theta _{12}&=&\frac{\left( -\frac{3\theta ^{2}}{8}+\frac{\theta }{2%
\sqrt{2}}+\frac{1}{2}\right) ^{2}+\left( -\frac{1}{8}\sqrt{3}\theta ^{2}-%
\frac{\theta }{2\sqrt{6}}-\frac{1}{2\sqrt{3}}\right) ^{2}}{f\left( \theta
,\psi ,\phi \right) },\hspace{20mm} \\[0.12in]
\sin ^{2}\theta _{13}&=&\left( \left( -\frac{1}{8}\sqrt{3}\theta ^{2}-\frac{%
\theta }{2\sqrt{6}}-\frac{1}{2\sqrt{3}}\right) \cos (\psi )+\left( \frac{%
\sqrt{3}\theta ^{2}}{4}-\sqrt{\frac{2}{3}}\theta +\frac{1}{\sqrt{3}}\right)
\sin (\psi )\cos (\phi )\right) ^{2}  \notag \\
&+&\left( \left( \frac{3\theta ^{2}}{8}-\frac{\theta }{2\sqrt{2}}-\frac{1}{2}%
\right) \cos (\psi )-\left( \frac{\sqrt{3}\theta ^{2}}{4}-\sqrt{\frac{2}{3}}%
\theta +\frac{1}{\sqrt{3}}\right) \sin (\psi )\sin (\phi )\right) ^{2}, \\
\sin ^{2}\theta _{23}&=&\frac{\left( \left( \frac{\sqrt{3}\theta ^{2}}{8}-%
\frac{1}{2}\sqrt{\frac{3}{2}}\theta -\frac{1}{2\sqrt{3}}\right) \cos (\psi
)+\left( \frac{1}{\sqrt{3}}-\frac{\sqrt{3}\theta ^{2}}{4}\right) \sin (\psi
)\cos (\phi )\right) ^{2}}{f\left( \theta ,\psi ,\phi \right) }  \notag \\
&+&\frac{\left( \left( \frac{3\theta ^{2}}{8}-\frac{\theta }{2\sqrt{2}}+%
\frac{1}{2}\right) \cos (\psi )-\left( \frac{1}{\sqrt{3}}-\frac{\sqrt{3}%
\theta ^{2}}{4}\right) \sin (\psi )\sin (\phi )\right) ^{2}}{f\left( \theta
,\psi ,\phi \right) }, \\
f\left( \theta ,\psi ,\phi \right) &=&-\left\{ \left( -\frac{1}{8}\sqrt{3}%
\theta ^{2}-\frac{\theta }{2\sqrt{6}}-\frac{1}{2\sqrt{3}}\right) \cos \psi
+\left( \frac{\sqrt{3}\theta ^{2}}{4}-\sqrt{\frac{2}{3}}\theta +\frac{1}{%
\sqrt{3}}\right) \sin \psi \cos \phi \right\} ^{2}  \notag \\
&-&\left\{ \left( \frac{3\theta ^{2}}{8}-\frac{\theta }{2\sqrt{2}}-\frac{1}{2%
}\right) \cos (\psi )-\left( \frac{\sqrt{3}\theta ^{2}}{4}-\sqrt{\frac{2}{3}}%
\theta +\frac{1}{\sqrt{3}}\right) \sin \psi \sin \phi \right\} ^{2}+1.
\label{e:fth}
\end{eqnarray}

\section{Computation of radiative decays $l \rightarrow l^{\prime}\protect\gamma$}
\label{a:llg}
The branching ratio of $l \rightarrow l^{\prime} \gamma$ is
\begin{align}
\text{Br}(l \rightarrow l^{\prime} \gamma) = \frac{\tau_{l} \alpha_{\text{EM}%
} m_l^5}{64 \pi^4} \left(|c_L|^2 + |c_R|^2\right),
\end{align}
where the Wilson coefficients $c_{L,R}$ are calculated up to order two. Because of the strong hierarchy in the Yukawa couplings in our model $|y_{l^{\prime}l}| \gg |y_{ll^{\prime}}|$ the contributions from $c_L$ to $\text{Br}(l\rightarrow l^{\prime} \gamma)$ can be neglected. The one-loop expressions corresponding to the diagrams in Fig. \ref{fig:diags} are given by \cite{Harnik:2012pb}
\begin{equation} \label{e:cr}
c_R^{\text{1Loop}} \simeq \frac{1}{4 m_l} \int^1_0 \delta (1-u-v-w)
\frac{v\,w\,m_{l}\,y_{l^{\prime}l}\,y^{\ast}_{ll} + (u+v)\,m_l\,y_{l^{\prime}l}\,y_{ll}}{w\,m_s^2 - v\,w\,m_l^2 + (u+v)\,m_l^2 - u\,v\,q^2} du\,dv\,dw
\end{equation}
with $s=h,H,\eta_I,\phi_{a,R},\phi_{a,I}$. Specifically in the case of $\mu \rightarrow e \gamma$ the two-loop contributions with a top quark and a $W$ boson running in the loop have to be taken into account as they can dominate the cross section in certain regions of the parameter space \cite{Harnik:2012pb,Chang:1993kw}. The top-loop analytical expressions are
\begin{align} \label{e:crtop}
c_R^{t\gamma}&=-\frac{8}{3}\kappa \frac{v}{m_t} y_{\mu\tau}\left[\text{Re}(y_{tt})f(z_{ts})-i\text{Im}(y_{tt})g(z_{ts})\right], \\
c_R^{tZ}&=-4\kappa \frac{(1-4 s_{\theta_W}^2)(1-\frac{8}{3} s_{\theta_W}^2) v}{16 s_{\theta_W}^2 c_{\theta_W}^2} y_{\mu\tau}\left[\text{Re}(y_{tt})\tilde{f}(z_{ts})-i\text{Im}(y_{tt})\tilde{g}(z_{ts})\right]
\end{align}
with $\theta_W \simeq 28.74^{\circ}$ and
\begin{align}
f(z) = &\frac{z}{2} \int_0^1 \frac{1-2x(1-x)}{x(1-x)-z}\log\frac{x(1-x)}{z}dx,\\
g(z)= &\frac{z}{2} \int_0^1 \frac{1}{x(1-x)-z}\log\frac{x(1-x)}{z} dx, \\
h(z)= &\frac{z}{2} \int_0^1 \frac{1}{z-x(1-x)}\left[1+ \frac{z}{z-x(1-x)}\log\frac{x(1-x)}{z}\right] dx, \\
\tilde{f}(x,y)=& \frac{y f(x)}{y-x} + \frac{x f(y)}{x-y}, \qquad \tilde{g}(x,y)= \frac{y g(x)}{y-x} + \frac{x g(y)}{x-y},\\ 
z_{ij} =& \frac{m_i^2}{m_j^2} \quad \text{and} \quad \kappa = \frac{\alpha}{2\sqrt{2}\pi} G_F \frac{v}{m_{l}}.
\end{align}
The $W$-loop expressions on the other hand are
\begin{align} 
c_R^{W\gamma}&=\kappa\,y_{\mu\tau}\left[3 f(z_{Ws}) + (5+\frac{3}{4}) g(z_{Ws}) + \frac{3}{4}) h(z_{Ws}) + \frac{f(z_{Ws})-g(z_{Ws})}{2 z_{Ws}} \right], \\ \label{e:crw}
c_R^{WZ}&=\kappa \frac{1-4 s_{\theta_W}^2}{4 s_{\theta_W}^2} y_{\mu\tau} \left[\frac{1}{2}(5-t_{\theta_W}^2) \tilde{f}(z_{ts}, z_{WZ}) + \frac{1}{2}(7-3t_{\theta_W}^2) \tilde{g}(z_{ts}, z_{WZ})\right. \\ \nonumber 
&\qquad + \left. \frac{3}{4} g(z_{ts}) + \frac{3}{4} h(z_{ts}) + \frac{1}{4 z_{ts}} (1-t_{\theta_W}^2) (\tilde{f}(z_{ts}, z_{WZ}) - \tilde{g}(z_{ts}, z_{WZ})) \right].
\end{align}
The corresponding diagrams of these loop contributions can be found in, e.g., Fig. 12 of \cite{Harnik:2012pb}.

\section{Loop factors for $h\rightarrow \protect\gamma \protect\gamma$}
\label{a:hgg}
The dimensionless loop factors $F_{1/2}\left( \beta \right) $ and $%
F_{1}\left( \beta \right) $ (for spin-$1/2$ and spin-$1$ particles in the
loop, respectively) appearing in Eq. (\ref{e:hggcr}) are given by \cite{Gunion:1989we,Djouadi:2005gj} 
\begin{align} \label{e:f12}
F_{1/2}\left( \beta \right) &=2\left[ \beta +\left( \beta -1\right) f\left(
\beta \right) \right] \beta ^{-2}, \\
F_{1}\left( \beta \right) &=-\left[ 2\beta ^{2}+3\beta +3\left( 2\beta
-1\right) f\left( \beta \right) \right] \beta ^{-2},  \label{F} \\
F_{0}\left( \beta \right) &=-\left[ \beta -f\left( \beta \right) \right]
\beta ^{-2},
\end{align}%
with 
\begin{equation} \label{e:fcase}
f\left( \beta \right) =%
\begin{cases}
\arcsin ^{2}\sqrt{\beta },\hspace{0.5cm}\mathit{for}\hspace{0.2cm}\beta \leq
1 \\ 
-\frac{1}{4}\left[ \ln \left( \frac{1+\sqrt{1-\beta ^{-1}}}{1-\sqrt{1-\beta
^{-1}}}\right) -i\pi \right] ^{2},\hspace{0.5cm}\mathit{for}\hspace{0.2cm}%
\beta >1.%
\end{cases}%
\end{equation}

\end{appendix}

\newpage

\bibliographystyle{h-physrev}
\bibliography{s4pap}

\end{document}